\documentclass[journal]{IEEEtran}
\usepackage{cite}
\usepackage{amsmath,amssymb,amsfonts,graphicx,subcaption}
\usepackage{algorithmic}
\usepackage{textcomp}
\usepackage{booktabs}
\usepackage{import}
\usepackage{multirow}
\usepackage{textcomp}
\usepackage{xcolor}
\usepackage{comment}
\usepackage[normalem]{ulem}
\usepackage{placeins}
\graphicspath{{images/}}

\newcommand\scalemath[2]{\scalebox{#1}{\mbox{\ensuremath{\displaystyle #2}}}}

\newcommand{\ca}[1]{\textcolor{blue}{#1}}

\def\BibTeX{{\rm B\kern-.05em{\sc i\kern-.025em b}\kern-.08em
    T\kern-.1667em\lower.7ex\hbox{E}\kern-.125emX}}

\title{Deep learning network to correct axial and coronal eye motion in 3D OCT retinal imaging} 
\author{Yiqian~Wang, 
        Alexandra~Warter,
        Melina~Cavichini,
        Varsha~Alex,
        Dirk-Uwe~G.~Bartsch,
        William~R.~Freeman,
        Truong~Q.~Nguyen,~\IEEEmembership{Fellow,~IEEE,}
        and~Cheolhong~An
\thanks{This work is supported in part by the National Eye Institute under grant 1R01EY033847. }
\thanks{Yiqian Wang, Truong Q. Nguyen, and Cheolhong An are with the Department
of Electrical and Computer Engineering, University of California, San Diego,
CA 92093, USA. e-mail: chan@eng.ucsd.edu}
\thanks{Alexandra Warter, Melina Cavichini, Varsha Alex, Dirk-Uwe G. Bartsch, and William R. Freeman are with the Jacobs Retina Center, Shiley Eye Institute, University of California, San Diego, CA 92093, USA.}
}

\begin{document}
\maketitle
\begin{abstract}
Optical Coherence Tomography (OCT) is one of the most important retinal imaging techniques. However, involuntary motion artifacts still pose a major challenge in OCT imaging that compromises the quality of downstream analysis, such as retinal layer segmentation and OCT Angiography. We propose deep learning based neural networks to correct axial and coronal motion artifacts in OCT based on a single volumetric scan. The proposed method consists of two fully-convolutional neural networks that predict Z and X dimensional displacement maps sequentially in two stages. The experimental result shows that the proposed method can effectively correct motion artifacts and achieve smaller errors than other methods. Specifically, the method can recover the overall curvature of the retina, and can be generalized well to various diseases and resolutions. 
\end{abstract}

\section{Introduction}
\label{sec:octmoco_intro}
\begin{figure}[ht] 
    \centering
    \includegraphics[width = 0.7\linewidth]{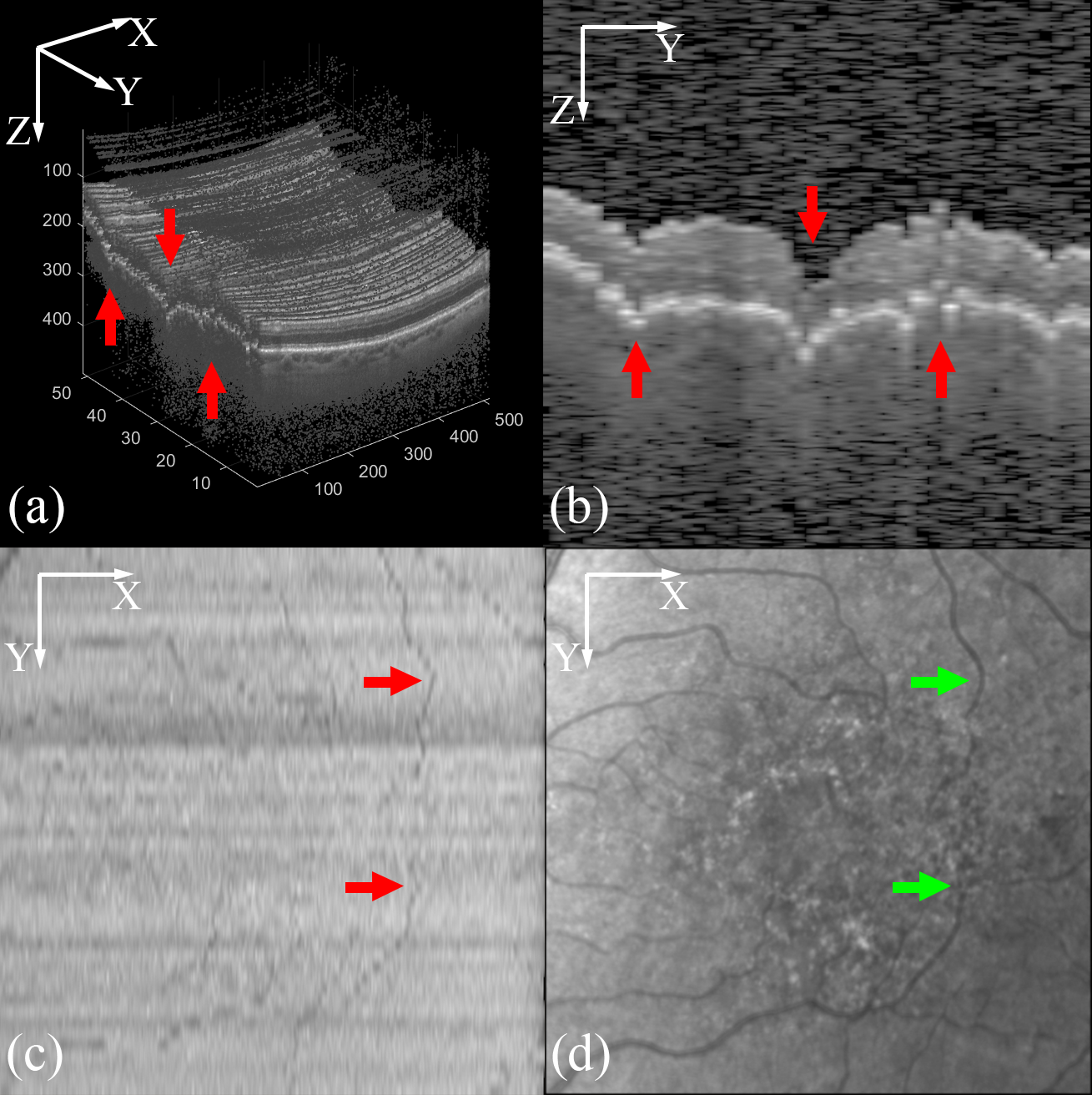}
    \caption{(a) The axial motion artifacts in 3D OCT volume indicated with red arrows, (b) cross-sectional B-scan (YZ plane) with motion artifacts, (c) en-face C-scan (XY plane) with motion artifacts, (d) IR en-face image reference with true vessel shapes indicated by green arrows. Gamma correction at 2.2 is applied on the OCT images for better visualization.}
    \label{fig:thumbnail}
\end{figure}

Optical Coherence Tomography (OCT) is a non-invasive imaging technique that visualizes cross-sectional images of biological tissues at micrometer-resolution \cite{huang1991optical}. The impact of OCT in retinal imaging is very significant in opthalmology, so that OCT has become the standard of care for diagnosing and monitoring most retinal diseases \cite{abramoff2010retinal}, including age-related macular degeneration (AMD), diabetic macular edema (DME), glaucoma, and so on. 
The imaging principle of OCT is based on low-coherence interferometry. The object is probed with low-coherent infrared light, and the depth of backscattered light along the beam axis is measured by interference. The interferogram intensities represent 1D depth (A-scan, Z axis of Fig.~\ref{fig:thumbnail})  from the backscattering. 2D cross-sectional images (B-scan, XZ plane of Fig.~\ref{fig:thumbnail}) are acquired in a sequence by moving the infrared beam through the object in a raster-scanning pattern. Finally, a 3D volume can be formed by stacking the B-scans (XZ planes) to the Y axis, as illustrated in Fig.~\ref{fig:thumbnail}. The direction for B-scan acquisition (X axis of Fig.~\ref{fig:thumbnail}) is called the \textit{fast scanning axis}, while the direction for stacking B-scans (Y axis of Fig.~\ref{fig:thumbnail}) is called the \textit{slow scanning axis}, and the plane spanned by the other two axes is called the \textit{coronal} or \textit{en-face} plane. 

Motion correction is one of the major challenges in OCT imaging, as motion artifacts would not only influence the visualization of the volumetric data, but also reduce the reliability of retinal biomarkers \cite{sanchez2019review}. Moreover, they may increase the difficulty of downstream tasks including disease classification and segmentation, layer segmentation and OCT-Angiography (OCT-A) imaging. Motion artifacts in OCT can be caused by head motion, respiration, vascular pulsation, and involuntary fixational eye movements. Specifically, even when the patient is instructed to stay still and fixate on an object, the eye may still carry out microsaccades, drifts, and tremors with various frequency and magnitude \cite{sanchez2019review}. These involuntary motions lead to axial and coronal distortions, shown in sub-images~(b) and (c) in Fig \ref{fig:thumbnail}, respectively, where their motion artifacts are indicated by red arrows. The axial motion introduces discontinuity in the cross-sectional B-scan as in sub-image (b), while coronal motion causes distortion of vessels in the en-face plane shown in sub-image (c) compared with the reference IR image in sub-image (d) in Fig. \ref{fig:thumbnail}.

Existing literature on OCT motion correction can be categorized into \textit{prospective} and \textit{retrospective} approaches \cite{sanchez2019review}. Prospective approaches include hardware eye-tracking systems that detect and compensate for motions during image acquisition \cite{ferguson2004tracking, tao2010interlaced}. Although hardware-based methods can usually achieve more accurate results, they are not available for every OCT device and their solution cannot eliminate all the motion artifacts. Retrospective approaches are applied after image acquisition and most are software-based methods. Many successful software methods require more than one OCT volume \cite{potsaid2008ultrahigh,kraus2012motion} or multimodal images as a reference, which introduces extra burden in clinical setting. Other methods based on a single OCT volume tend to remove the curvature of the retina generating overly-smoothed result \cite{antony2011automated, xu2012alignment}. Moreover, the macular curvature needs to be preserved as it is used in detecting diseases including myopic \cite{gaucher2008dome, park2019influence}, as well as normal eyes \cite{minami2020analysis}.

In this paper, we propose a deep learning method that utilizes fully convolutional neural networks to correct both axial and coronal motion artifacts in OCT with a single input volume. The axial motion correction network is able to predict axial displacement and recover the overall curvature of the retina. The coronal motion correction network extracts vessel segmentation information and compensates for the remaining error of eye-tracking hardware in the fast-scanning direction.  
To the best of our knowledge, the proposed motion correction network is the first deep learning approach for the OCT motion correction problem, and achieves significant improvement upon the conventional methods from the experimental results, while achieving the least distortion to retinal curvature. Moreover, the proposed method can be generalized well to various diseases and resolutions.

\section{Related work}
\label{sec:related}
Although artifacts caused by involuntary eye motion is a fundamental problem in retinal OCT, it has not been widely researched according to existing literature. Existing literature in OCT motion correction have been extensively reviewed and summarized in two papers published in 2017 \cite{baghaie2017involuntary} and 2019 \cite{sanchez2019review}.
Most works correct axial and coronal movement by treating fast B-scans as artifact-free rigid bodies, since the acquisition speed of a fast B-scans is faster than that of the expected eye motion using modern OCT devices \cite{sanchez2019review}. Axial movement is observed to be more significant than coronal movement in magnitude \cite{potsaid2008ultrahigh}, and the axial resolution is also often higher than coronal resolution \cite{potsaid2008ultrahigh, kraus2012motion}. 
Coronal artifacts are caused by eye movement in the 2D en-face plane. The X component of such motion, also called \textit{in-plane} motion, is parallel to the X-Z plane of B-scans and can be observed by discontinuities in retinal vessels, which are prominent features in the en-face plane. The Y motion along the slow scanning axis, also called \textit{out-of-plane} motion, is most difficult to quantify \cite{baghaie2017involuntary}. Negative or positive displacement to the Y axis can cause repeated B-scans of the same region or larger gaps between neighboring B-scans. 

Most existing approaches can be divided in two major categories: prospective and retrospective approaches \cite{sanchez2019review}. 
Prospective approaches often include active eye-tracking hardwares mounted on OCT devices \cite{ferguson2004tracking, tao2010interlaced} and usually produce more accurate alignment results \cite{baghaie2017involuntary}.
There are also approaches that depend on specially designed scanning patterns \cite{chen2017three} or signal acquisition techniques \cite{ksenofontov2020numerical} to obtain an artifact-free OCT volume. Nevertheless, they are difficult to implement in existing OCT systems and cannot correct eye movements in conventional OCT scans.
The retrospective approaches on the other hand are software-based solutions.
Many motion correction algorithms require more than one OCT volume, either repeated in the same direction \cite{gibson2010optic, niemeijer2012registration}, or both horizontal and vertical scans in orthogonal directions \cite{potsaid2008ultrahigh, kraus2012motion}.
Potsaid et al. applied orthogonal OCT volumes in both horizontal and vertical directions that corrects the axial motion \cite{potsaid2008ultrahigh}. 
Gibson et al. proposed an axial motion correction algorithm based on optic nerve head segmentation surface that requires parallel OCT volumes scanned in the same direction  \cite{gibson2010optic}. 
Niemeijer et al. proposed a graph-based method to register multiple OCT volumes and find the optimal translation for each A-scan that can correct both X and Z motion \cite{niemeijer2012registration}.
Wu et al. proposed a registration method for 3D OCT volumes based on Coherent Point Drift \cite{wu2014stable}, which includes a motion correction algorithm that detects and deforms 3D vessel center lines in source and target volumes.
The method proposed by Kraus et al. \cite{kraus2012motion, kraus2014quantitative} registers two or more OCT volumes in orthogonal directions, which sequentially corrects axial and coronal motion in two stages. It has been widely adopted as a standard pre-processing algorithm in OCT-A imaging \cite{sanchez2019review}.
However, these methods need to capture multiple OCT volumes, such that they increase (double) the time required for clinical examinations and impose additional burden on limited medical resources.

Other methods for estimating eye motion based on a single OCT volume are proposed to save imaging time. 
The method proposed by Antony et al. \cite{antony2011automated} utilizes segmentation of the retinal pigment epithelium (RPE) layer and thin-plate spline fitting. A major drawback of the method is that it flattens the RPE layer into a plane, which leads to artifacts for diseases that manifest in the RPE layer. It is also undesirable to observe diseases including myopic \cite{gaucher2008dome, park2019influence} as it removes the curvature of the retina.
Xu et al. \cite{xu2012alignment} proposed a particle filter method to correct axial and X directional motion in optic nerve head (ONH) centered OCT volumes. The method also results in flattened RPE surface, and is only validated on synthetic motions within 2-10 pixel range.
The algorithm of Montuoro et al. \cite{montuoro2014motion} corrects axial motion by smoothing the RPE segmentation with a local symmetry assumption, and accounts for X directional coronal motion based on phase shift in the Fourier domain. It can recover the retinal curvature to some extend, but the symmetry assumption does not always hold for retina with diseases.
Fu et al. \cite{fu2016eye} proposed a method to correct both axial and X directional motion based on saliency detection, but the authors only tested the performance of the X motion correction using synthetic data with X motion smaller than 15 pixels.

In our previous work \cite{wang2021icip}, we applied the deep learning algorithm to correct the axial motion. It improved correction accuracy while preserving retinal curvature. In this paper, we propose the deep learning network to jointly correct both coronal and axial motions. We include a comprehensive comparison with several methods using different metrics for OCT volumes with analysis of various diseases and different resolutions. The preservation of curvature is explicitly evaluated by the curvature and distortion coefficient. Ablation studies of segmentation input, post-processing, and stand-alone coronal motion correction network are included. We also add qualitative comparison of the layer segmentation and vessel segmentation using different motion correction methods.

\section{Axial motion correction network}

We first apply axial motion correction to eliminate large Z displacement before correcting coronal motion, because the axial motion is more significant compared with coronal motion in retinal OCT. It has been observed that the axial motion is larger in micrometers compared with coronal motion \cite{potsaid2008ultrahigh, kraus2012motion}, and the fact that the axial direction has higher resolution in most OCT systems \cite{potsaid2008ultrahigh, rocholz2018spectralis, huang2016optical} makes the axial artifacts more dominant.

\label{sec:octmoco_axial}
\subsection{Network architecture}

\begin{figure*}[ht]
    \centering
    \includegraphics[width = 0.8\linewidth]{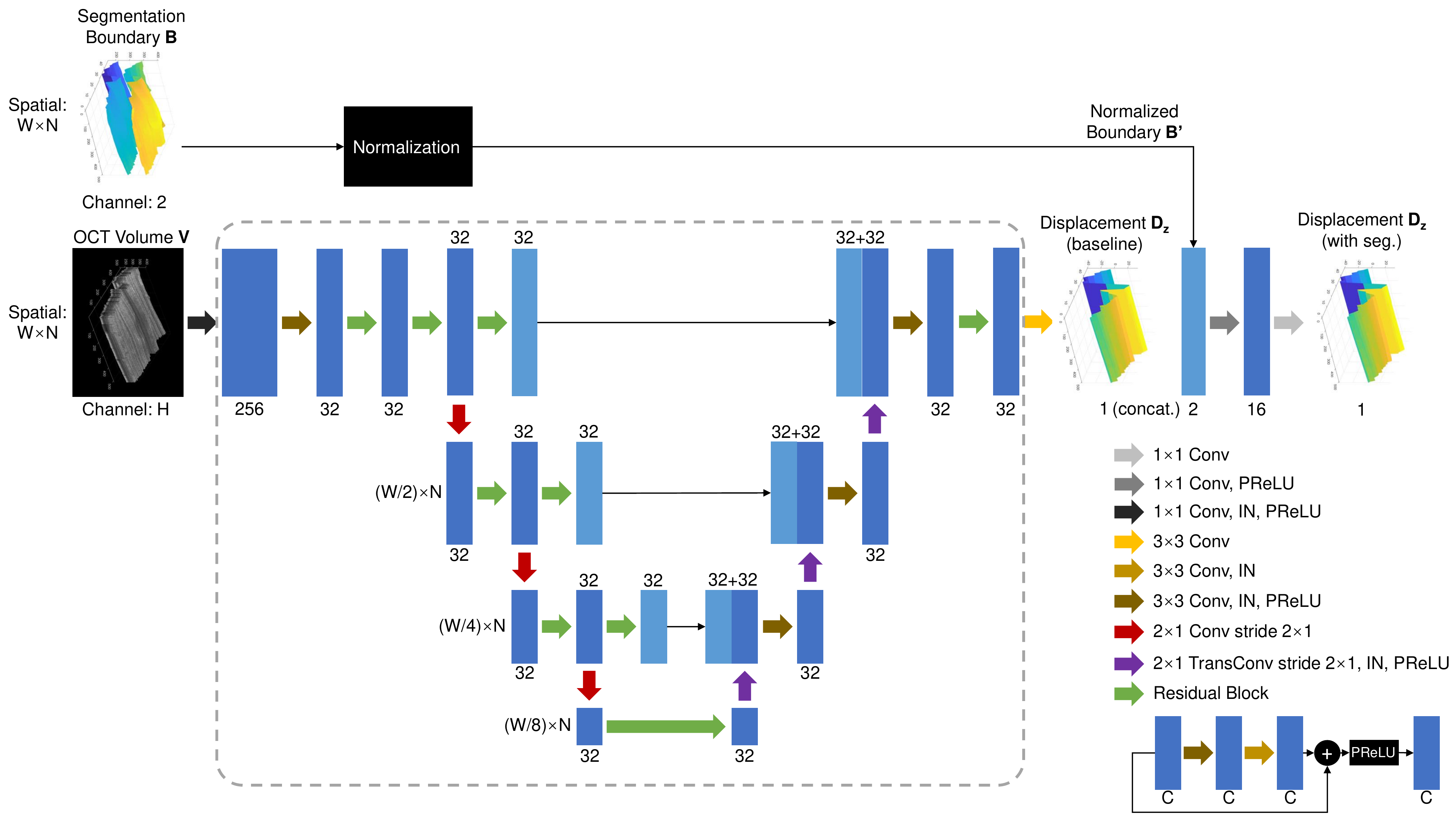}
    \caption{Architecture of the proposed OCT motion correction network to predict an axial displacement map. The baseline network is circled in dashed lines, and the network with segmentation concatenates the baseline output with the normalized segmentation boundary $\mathbf{B}'$ to enhance the final displacement prediction. }
    \label{fig:network}
\end{figure*}

We present a modified U-Net \cite{ronneberger2015unet} with residual blocks to predict a displacement map for a single OCT volumetric input. 
The proposed network takes any number of stacked B-scans due to the fully convolutional architecture.
The proposed baseline method of Fig.~\ref{fig:network} operates on a single OCT volumetric scan $\mathbf{V}\in\mathbb{R}^{H\times W\times N}$ where $W$ and $H$ are the width and height of each B-scan and $N$ is the number of B-scans. Z axis of the input OCT volume is treated as channels ($H$), while X and Y axes are considered as spatial dimensions ($W \times N$). 
The network outputs a displacement map $\mathbf{D}_z\in\mathbb{R}^{W\times N}$ where each pixel contains a displacement value to Z axis.
Negative displacement shifts the A-scan upwards and positive displacement shifts it downwards. The magnitude of displacement denotes the number of pixels to be shifted and it is divided by a normalization factor $Z_\mathrm{norm}$ , which is a tunable hyper-parameter to scale predicted Z displacement, for better numerical stability. Finally, the motion corrected OCT volume $\mathbf{V}_\mathrm{dz}$ is derived for a given displacement $\mathbf{D}_z$ {in floating point value} as 
\begin{equation}
    \mathbf{V}_\mathrm{dz}(z,x,y) = \mathbf{V}(z-\mathrm{int}(Z_\mathrm{norm}\mathbf{D}_z(x,y)),~x,~y),
\end{equation}
{where $(x,y,z)\in[0,W-1]\times[0,N-1]\times[0,H-1]$ and $\mathrm{int}(\cdot)$ denotes integer conversion.}

As illustrated in Fig.~\ref{fig:network}, the network includes 4 resolution levels similar to U-Net. 
$1\times 1$ convolution is applied at the first layer to compress the number of channels, and it is followed by $3\times 3$ convolutions for further processing. Instance normalization (IN) is applied after convolutions in order to normalize over the spatial dimensions without being influenced by other volumes in the same batch. 
The skip connection is removed at the original resolution to enhance smoothness of prediction and reduces memory consumption.
For the three downsampling blocks denoted by red arrows, $2\times 1$ convolution with stride $2\times 1$ is adopted to downsample the X dimension by 2, while keeping the resolution on the Y dimension unchanged, since the number of B-scans $N$ in our dataset is significantly smaller than the width of B-scans $W$. Similarly, $2\times 1$ transposed convolution with stride $2\times 1$ is used to upsample X dimension by 2 in three upsampling blocks which are indicated by purple arrows. Differently from the original U-Net, the input features on the same resolution are processed by residual blocks before concatenation with the upsampled ones. Dropouts are applied at blue blocks with black dashed contour in Fig.~\ref{fig:network} to prevent overfitting during training.

We also propose an enhanced version of the baseline architecture by including the segmentation of the inner limiting membrane (ILM) and the retinal pigment epithelium (RPE) layer. As shown in Fig.~\ref{fig:network}, we first normalize two segmentation boundaries and concatenate them with the output of the network, and then apply two additional layers with $1\times 1$ convolution to get the displacement prediction. The boundary normalization is computed by the following steps. 
{We denote the two segmentation boundaries with $\mathbf{B}\in\mathbb{R}^{2\times W\times N}$, where $\mathbf{B}(0,x,y)$ and $\mathbf{B}(1,x,y)$ entries represent Z coordinates of ILM and RPE layers at pixel $(x,y)$, respectively. }
The overall retinal {tilt} $\mathbf{T}\in \mathbb{R}^{2\times W\times N}$ is first computed by 
\begin{equation}
    \mathbf{T}(z,x,y) = \mathbf{B}(z,x,0)+\frac{\mathbf{B}(z,x,N-1)-\mathbf{B}(z,x,0)}{N-1}y,
\end{equation}
where $z \in \{0,1\}$ and $(x,y) \in [0,W-1]\times [0, N-1]$.
Then, the normalized boundaries $\mathbf{B}'$ can be obtained by
\begin{equation}
    \mathbf{B}' = (\mathbf{T} - \mathbf{B})/Z_\mathrm{norm}.
\end{equation}

\subsection{Ground truth acquisition}
\begin{figure}[htb]
    \centering
    \includegraphics[width = 0.95\linewidth]{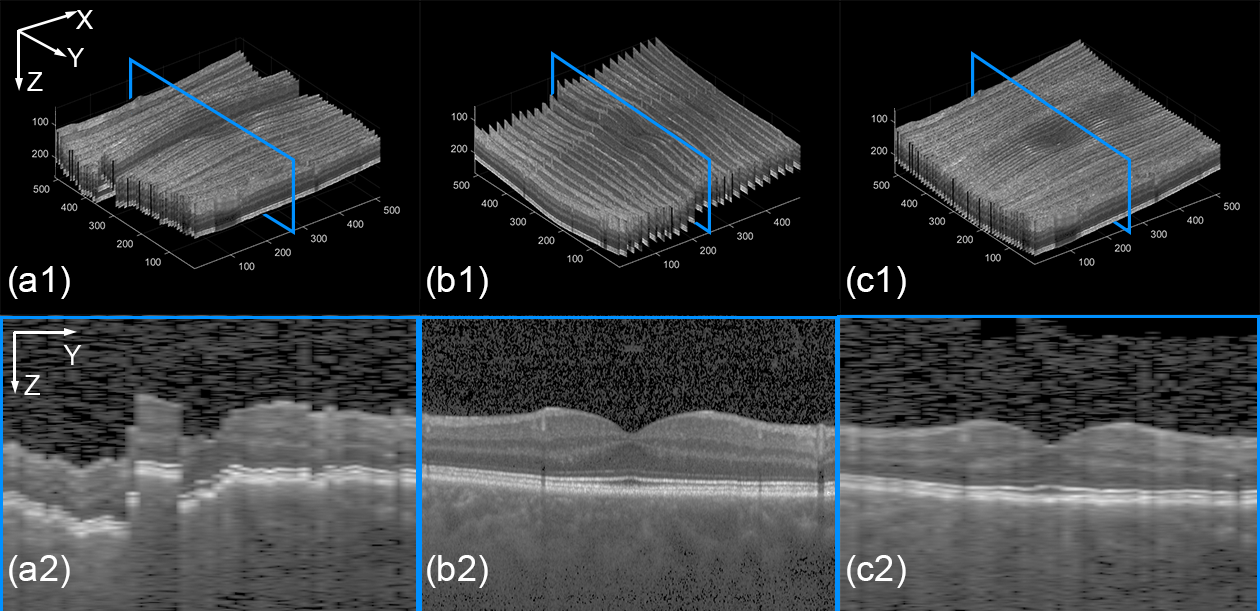}
    \caption[Orthogonal method for ground truth acquisition. Column (a) shows the horizontal volume, (b) shows the paired vertical volume, (c) shows the motion-corrected horizontal volume using the motion correction algorithm. Row (1) shows the 3D volumes, and row (2) shows the cross-sectional B-scan.]
    {Orthogonal method \cite{potsaid2008ultrahigh} for ground truth acquisition. Column (a) shows the horizontal volume, (b) shows the paired vertical volume, (c) shows the motion-corrected horizontal volume using the motion correction algorithm in \cite{potsaid2008ultrahigh}. Row (1) shows the 3D volumes, and row (2) shows the cross-sectional B-scan. }
    \label{fig:GT_method}
\end{figure}

In order to obtain ground truth (motion artifacts-free) volumes and corresponding displacement maps, pairs of horizontal and vertical 3D OCT volumes with motion artifacts are collected, and each volume is corrected with its orthogonal reference using the motion correction algorithm in \cite{potsaid2008ultrahigh}, as illustrated in Fig.~\ref{fig:GT_method}.
Note that we use horizontal and vertical volume pairs for ground truth, but the proposed network takes only one single (horizontal or vertical) 3D OCT volume as input.

\subsection{Post processing with linear least squares}
During the inference time, we apply a post-processing step where a linear function is fitted to the X axis of the predicted displacement $\mathbf{D}_z$ via linear least squares, in order to guarantee that the resulting fast B-scans in $\mathbf{V}_\mathrm{dz}$ have linear displacement in the Z direction. 
Specifically, denote the coordinates $\mathbf{X}\in\mathbb{R}^{W\times 2}$ as
\begin{equation}
    \mathbf{X}=\begin{bmatrix}
    0,&1,&\cdots,&W-1 \\
    1,&1,&\cdots,& 1
    \end{bmatrix}^T, 
    \label{eq:octmoco_xcoord}
\end{equation} 
the line parameters $\boldsymbol{\beta} \in \mathbb{R}^{2\times N}$ can be obtained by solving the linear least squares problem
\begin{equation}
    \boldsymbol{\beta}^*(y) = \underset{\beta(y) \in \mathbb{R}^{2\times 1}}{\arg\min} \Big\| \mathbf{D}_z(y)-\mathbf{X}\beta(y) \Big\|^2.
\end{equation}

where $\boldsymbol\beta(y)$ and $\mathbf{D}_z(y)$ denote the $y$-th row of $\boldsymbol{\beta}$ and $\mathbf{D}_z$, respectively.
The solution is derived as
\begin{equation}
    \boldsymbol{\beta}(y) = (\mathbf{X}^T\mathbf{X})^{-1}\mathbf{X}^T\mathbf{D}_z(y) ,\quad y=0,\cdots, N-1
\end{equation}
and the displacement map obtained by solving the weighted least squares problem would be
\begin{equation}
    \mathbf{D}_z'(y) = \mathbf{X}\boldsymbol{\beta}(y)= \mathbf{X}(\mathbf{X}^T\mathbf{X})^{-1}\mathbf{X}^T\mathbf{D}(y) ,\quad y=0,\cdots, N-1.
    \label{eq:Dzp}
\end{equation}

\begin{figure}[htb]
    \centering
        \begin{subfigure}[b]{0.4\linewidth}
          \includegraphics[width = 0.8\linewidth]{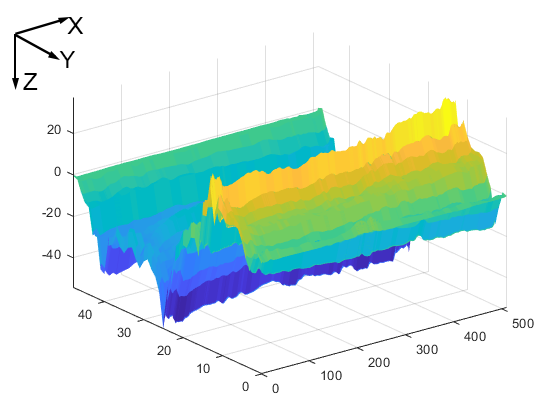}
          \caption{Without post-processing}
          \label{subfig:no_post}
        \end{subfigure}
        \begin{subfigure}[b]{0.4\linewidth}
          \includegraphics[width = 0.8\linewidth]{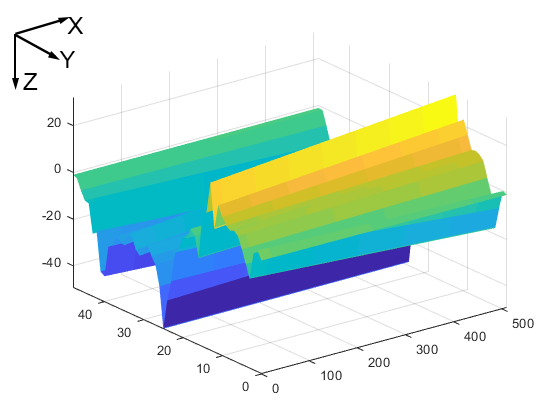}
          \caption{With post-processing}
          \label{subfig:post}
        \end{subfigure}
    \caption{Post processing with linear least squares. (a) shows the displacement without post-processing, (b) shows the displacement after post-processing.}
    \label{fig:ls_post}
\end{figure}

Fig.~\ref{fig:ls_post} illustrates an example, where the displacement maps without post-processing and with post-processing are shown in (a) and (b), respectively. It can be observed that the noise in (a) is removed along the X axis after the least square line fitting step.

\subsection{Loss function}
The loss function consists of two loss terms, including a displacement L1 loss, and a displacement smoothness loss. Denoting the predicted displacement $\mathbf{D}_z$ and the ground truth displacement $\mathbf{D}_z^\mathrm{GT}\in\mathbb{R}^{W\times N}$, the displacement L1 loss is given by
\begin{equation}
    \scalemath{0.9}{\mathcal{L}_\mathrm{disp}(\mathbf{D}_z;\mathbf{D}_z^\mathrm{GT}) = {\mathrm{mean}} \Big(\mathbf{M} \odot |\mathbf{D}_z - \mathbf{D}_z^\mathrm{GT}| \Big),}
\end{equation}
where $|\cdot |$ denotes the absolute value, $\odot$ is an element-wise multiplication, and $\mathbf{M}\in\mathbb{R}^{W\times N}$ is a predefined mask in [0,1], for assigning more weight at the center and less weight at the boundary of the OCT scan, as the center is the most important region of interest in clinical applications. 

The second term is a displacement smoothness loss inspired by \cite{yu2019robust} to enforce smoothness along the fast-scanning axis
\begin{equation}
    \mathcal{L}_\mathrm{smooth}(\mathbf{D}_z, \mathbf{D}_z') = {\mathrm{mean}} \left|\mathbf{D}_z - \mathbf{D}'_z\right|,
\end{equation}
which is an L1 error between the raw displacement $\mathbf{D}_z$ and the least squares smoothed $\mathbf{D}'_z$ in eq. (\ref{eq:Dzp}).
Finally, the total loss is a weighted sum of the two loss terms
\begin{equation}
    \mathcal{L} =  \lambda_\mathrm{disp}\mathcal{L}_\mathrm{disp} + \lambda_\mathrm{smooth} \mathcal{L}_\mathrm{smooth},
\end{equation}
where $\lambda_\mathrm{disp}$ and $\lambda_\mathrm{smooth}$ are tunable weighting parameters.

\subsection{Data augmentation}
\begin{figure}[htb]
    \centering
        \begin{subfigure}[b]{0.31\linewidth}
          \includegraphics[width = 1\linewidth]{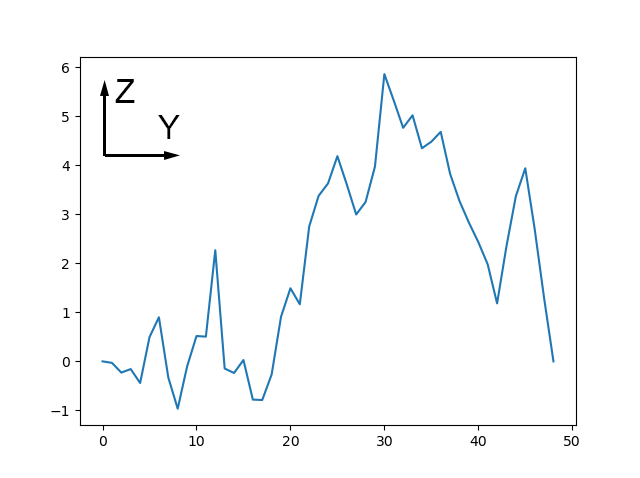}
          \caption{Y dimension $\boldsymbol{\delta}_\mathrm{Y}$}
          \label{subfig:aug_y}
        \end{subfigure}
          \hfill
        \begin{subfigure}[b]{0.31\linewidth}
          \includegraphics[width = 1\linewidth]{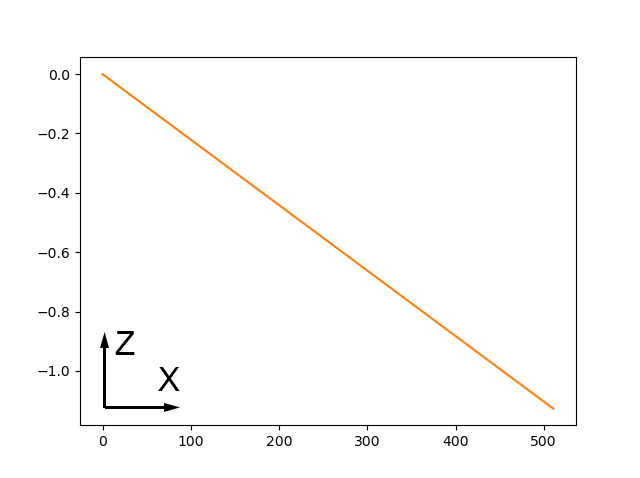}
          \caption{X dimension $\boldsymbol{\delta}_\mathrm{X}$}
          \label{subfig:aug_x}
        \end{subfigure}
          \hfill
        \begin{subfigure}[b]{0.31\linewidth}
          \includegraphics[width = 1\linewidth]{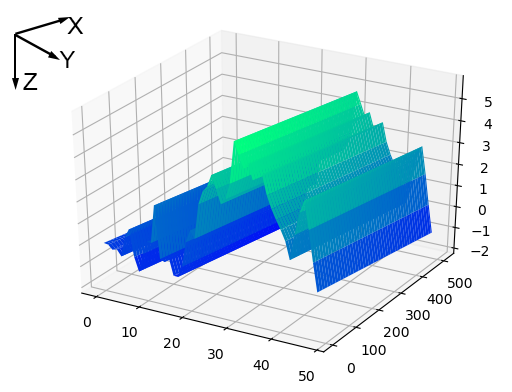}
          \caption{Total $\boldsymbol{\delta}$}
          \label{subfig:aug_disp}
        \end{subfigure}
    \caption{Data augmentation with random axial displacement. (a) Augmentation on Y dimension $\boldsymbol{\delta}_\mathrm{Y}$, (b) augmentation on X dimension $\boldsymbol{\delta}_\mathrm{X}$, (c) Total augmentation $\boldsymbol{\delta}$ is applied to the input OCT volume.}
    \label{fig:aug}
\end{figure}

The input OCT volume and segmentation boundaries in the training data are augmented by random flipping on X and Y axes to prevent over-fitting. We also add random displacement for augmentation on top of the existing eye motion as follows: An $N$-dimensional Gaussian random vector with zero mean and unit variance $\mathbf{g}=(g_0,g_1,\cdots,g_N)^T$ is generated, and its cumulated sum is computed as
\begin{equation}
    \begin{bmatrix}
    g_0, & g_0+g_1, & \cdots, & \sum_{k=0}^{N-1} g_k
    \end{bmatrix}^T.
\end{equation}
Then the tilt is removed from the axial augmentation on Y dimension $\boldsymbol{\delta}_\mathrm{Y}\in\mathbb{R}^{N\times 1}$ where $\boldsymbol{\delta}_\mathrm{Y}(0)=\boldsymbol{\delta}_\mathrm{Y}(N-1)=0$, and the $n$-th element is
\begin{equation}
    \boldsymbol{\delta}_\mathrm{Y}(n) = \sum_{k=1}^{n} g_k - \frac{n}{N-1}\sum_{k=1}^{N-1} g_k,
    ~~n=1,...,N-2
\end{equation}
and an example of $\boldsymbol{\delta}_\mathrm{Y}$ is shown in Fig.~\ref{fig:aug}(a). The axial augmentation on X dimension $\boldsymbol{\delta}_\mathrm{X} \in \mathbb{R}^{W\times 1}$ is generated by interpolating between 0 and a random number drawn from {a Gaussian random variable with zero mean and unit variance}, as shown Fig.~\ref{fig:aug}(b). Finally, the total augmentation $\boldsymbol{\delta} = \mathbf{1}_{W}\boldsymbol{\delta}_\mathrm{Y}^T + \boldsymbol{\delta}_\mathrm{X}\mathbf{1}_{N}^T$ is applied to the input OCT volume, where $\mathbf{1}_{k} \in \mathbb{R}^{k\times 1}$ denotes a $k$-dimensional vector of ones. Finally, $\boldsymbol{\delta}_\mathrm{Y}$ is subtracted from the ground truth displacement ${\mathbf{D}_z^\mathrm{GT}} = \mathbf{D}_z^\mathrm{GT} - \mathbf{1}_{W}\boldsymbol{\delta}_\mathrm{Y}^T$.

\section{Coronal motion correction network}
In the proposed coronal motion correction network, we only focus on X motion for two reasons. Firstly, Y motion is very small compared to distance between neighboring B-scans based on statistics in \cite{sanchez2019review}. Secondly, it is difficult to obtain ground truth of Y motion using conventional approaches, which will be discussed in subsection \ref{subsec:x_gt}.
The X motion correction network aims to predict a 1D displacement vector to the X axis {$\mathbf{D}_x\in \mathbb{R}^{N\times 1}$}, 
where $N$ is the number of B-scans. Negative displacement shifts the B-scan left and positive displacement shifts it right. The magnitude of displacement denotes the number of pixel to be shifted and it is  divided by a normalization factor $X_\mathrm{norm}$ for better numerical stability. Similar to the axial motion correction, the X motion corrected OCT volume $\mathbf{V}_\mathrm{dx}$ is derived by shifting the Z corrected volume $\mathbf{V}_\mathrm{dz}$ by $\mathbf{D}_x$ displacement to the X axis as 
\begin{equation}
    \mathbf{V}_\mathrm{dx}(z,x,y) = \mathbf{V}_\mathrm{dz} \left(z,~x-\mathrm{int}\left(X_\mathrm{norm}\mathbf{D}_x(y)\right),~y\right),
\end{equation}
where {$(x,y,z)\in[0,W-1]\times[0,N-1]\times[0,H-1] $} and $X_\mathrm{norm}$ is a normalization factor. 

\subsection{Ground truth acquisition}
\label{subsec:x_gt}
Many methods in the literature require registering and jointly optimizing multiple orthogonal or parallel OCT volumes \cite{niemeijer2012registration, kraus2012motion, kraus2014quantitative, wu2014stable}. 
Furthermore, since their methods use a large number of B-scans in each OCT volume (e.g. $496\times 512\times 512$) to obtain C-scans with high-resolution in both fast and slow scanning axes, they fail to achieve visually desirable performance when resolution of the slow axis is low, as in our dataset ($496\times 512\times 49$). Therefore, it is a major challenge to prepare the appropriate ground truth for training the coronal motion correction network.

The ground truth for X directional displacement $\mathbf{D}_x^\mathrm{GT}$ is obtained from the HEYEX software by Heidelberg Spectralis, which corrects residual X motion from the hardware eye-tracking system \cite{teussink2019spectralis}. The magnitude of extracted X displacement is on average 0.96 pixel and 5 pixels at the maximum. 

\subsection{Network design}

\begin{figure*}[htb]
    \centering
    \includegraphics[width = \linewidth]{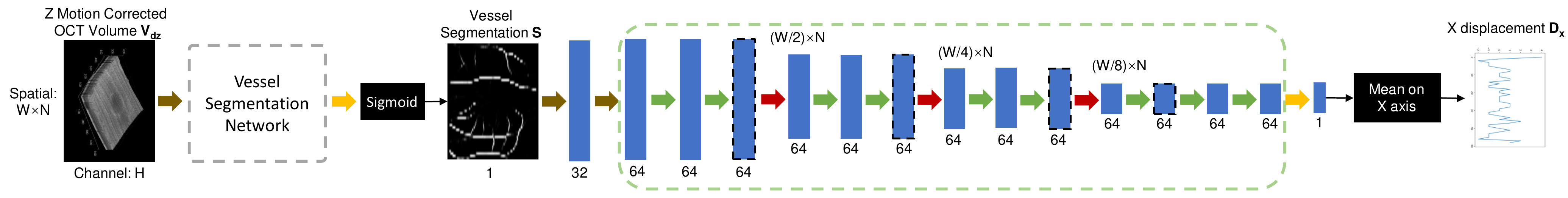}
    \caption{Network architecture of the vessel segmentation-based X motion correction network. {The vessel segmentation network in dashed block adopts a similar structure as the dashed block in Fig. \ref{fig:network}. } }
    \label{fig:xmoconet}
\end{figure*}

The network consists of a segmentation sub-network and a motion prediction sub-network. The segmentation sub-network first extracts a 2D vessel segmentation map on the en-face plane, where the coronal motion artifacts can be best observed. A subsequent motion prediction network then predicts a 1D displacement map $\mathbf{D}_x$ for motion to the X axis from the 2D vessel segmentation map.

\subsubsection{Vessel segmentation}
Since the X motion can be best observed by discontinuities of retinal vessel in OCT C-scans, and many related works also extract vessels for coronal motion correction \cite{niemeijer2012registration, wu2014stable}, we apply a vessel segmentation sub-network before the X motion prediction sub-network to extract critical information. 
Illustrated by gray dashed block in Fig.~\ref{fig:xmoconet}, the vessel segmentation network adopts a U-Net architecture similar to Fig. \ref{fig:network}, which takes the Z motion corrected volume $\mathbf{V}_\mathrm{dz}$ to obtain C-scan vessel segmentation $\mathbf{S}$ in the en-face plane.

\begin{figure}[htb]
    \centering
    \includegraphics[width = \linewidth]{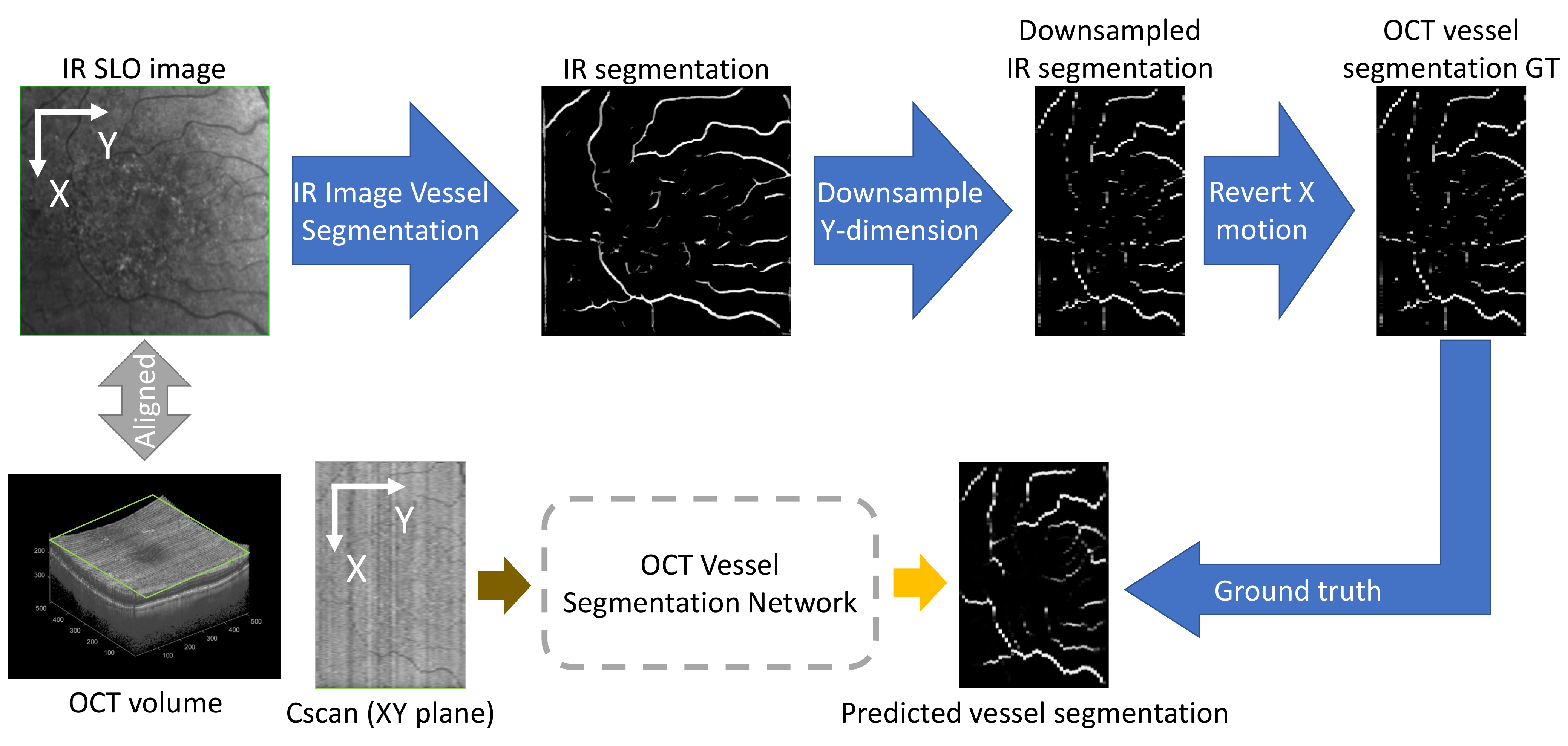}
    \caption{Procedure to obtain ground truth for training the OCT vessel segmentation sub-network.}
    \label{fig:x_training}
\end{figure}

In order to train the OCT vessel segmentation sub-network, we utilize the IR multimodal information captured concurrently with the OCT volume, which is aligned with the OCT en-face C-scan. In Fig. \ref{fig:x_training}, we apply the IR vessel segmentation network \cite{wang2021tip} to extract the probability of vessel (vesselness) from the IR image, and it is downsampled to $W\times N$ to match the resolution of OCT C-scan. The downsampled segmentation of the IR image is then converted to binary label by thresholding at 0.5 and used as ground truth $\mathbf{S}^\mathrm{GT}$ for training the OCT vessel segmentation sub-network. 

The segmentation sub-network is first trained with a combination of binary cross-entropy loss and soft Dice loss.
The binary cross-entropy loss can be expressed as
\begin{align}
\begin{split}
    \mathcal{L}_\mathrm{BCE}(\mathbf{S},\mathbf{S}^\mathrm{GT}) = -{\mathrm{mean}} \Big( \mathbf{S}^\mathrm{GT}\log\mathbf{S} \\ 
     +(1-\mathbf{S}^\mathrm{GT})\log(1-\mathbf{S}) \Big).
\end{split}
\end{align}
Note that the Sigmoid function for the predicted segmentation $\mathbf{S}$ can be integrated into the binary cross-entropy loss using the ``log-sum-exp“ trick for better numeric stability.
The soft Dice loss is defined as one minus the soft Dice coefficient
\begin{equation}
    \mathcal{L}_\mathrm{Dice}(\mathbf{S},\mathbf{S}^\mathrm{GT}) = 1-\frac{2\sum{\mathbf{S}\odot \mathbf{S}^\mathrm{GT}} }
    {\sum\mathbf{S}+ \sum \mathbf{S}^\mathrm{GT} },
\end{equation}
where $\odot$ denotes elementwise product, which measures the overlapping ratio between predicted vessel and ground truth.
Finally, the vessel segmentation sub-network is trained with a total loss and $\lambda_\mathrm{BCE}=\lambda_\mathrm{Dice}=1$, 
\begin{equation}
    \mathcal{L}_\mathrm{vessel} = \lambda_\mathrm{BCE}\mathcal{L}_\mathrm{BCE} + \lambda_\mathrm{Dice}\mathcal{L}_\mathrm{Dice}.
\end{equation}

\subsubsection{X motion prediction}
After training the vessel segmentation sub-network, we freeze the weights in the vessel segmentation network and train a X motion prediction sub-network to predict X displacement based on the 2D vessel segmentation map. The input to the X motion prediction sub-network is $1 \times W\times N$ with $B$ batch size. In the X motion prediction network, the resolution on the fast scanning axis is reduced by 3 downsampling convolutions with kernel size $2 \times 1$ and stride $2\times 1$.
The network output is synthesized by averaging across the fast scanning axis to obtain $\mathbf{D}_x$. Since the X displacement is a 1D vector whose element is a single displacement for each B-scan slice, we simplify the upsampling branch in the U-Net structure. 
The MSE Loss function is applied between predicted and ground truth displacement with normalization factor $X_\mathrm{norm}$:
\begin{equation}
    \mathcal{L}_\mathrm{xdisp} = \underset{y}{\mathrm{mean}} ||X_\mathrm{norm}\mathbf{D}_x(y) - \mathbf{D}_x^\mathrm{GT}(y)||^2.
\end{equation}

\section{Experimental result}
In the experiment, we compare the axial and coronal motion correction performance of our proposed network to that of four other methods \cite{montuoro2014motion, antony2011automated, fu2016eye, teussink2019spectralis} which operate on a single OCT volume input.

\subsection{Dataset}
We evaluate the performance of motion correction algorithms on two datasets collected by Jacobs Retina Center. The first dataset consists of 99 eyes with paired horizontal and vertical OCT volumes which are obtained by Heidelberg Spectralis in an imaging volume of $1.9 \times 5.8 \times 5.8$ (mm$^3$) with 20 degree field of view. Hardware eye-tracking is always turned on during the imaging process. All the OCT scans come with instrument's segmentation boundaries of 11 retinal layers.
Among 99 horizontal and 99 vertical volumes, the dimensions of 9 volumes are $496 \times 512 \times 25$, while those of the remaining 189 volumes are $496 \times 512 \times 49$. The 198 OCT volumes (99 horizontal and 99 vertical) are divided into 142, 19, 37 for training, validation, and testing, respectively. 
The dataset include both healthy subjects as well as patients with wet and dry AMD, diabetic retinopathy, and other diseases including epi-retinal membrane, macular edema, retinal detachment, macular hole, chorioretinopathy, and posterior vitreous detachment.

The second dataset includes only 106 singular OCT volumes (horizontal direction only). This dataset is also captured with the Heidelberg Spectralis OCT system, but covers a wider range of resolutions: there are 95 OCT volumes of the same resolution $496\times 512 \times 49$ in the training dataset, while there are also 3 volumes of resolution $496\times 512 \times 25$, 6 volumes of resolution $496\times 1024 \times 97$, and 2 volumes of resolution $496\times 1024 \times 49$. 


\subsection{Tilt correction}
\begin{figure}[htb]
    \centering
    \includegraphics[width = 0.8\linewidth]{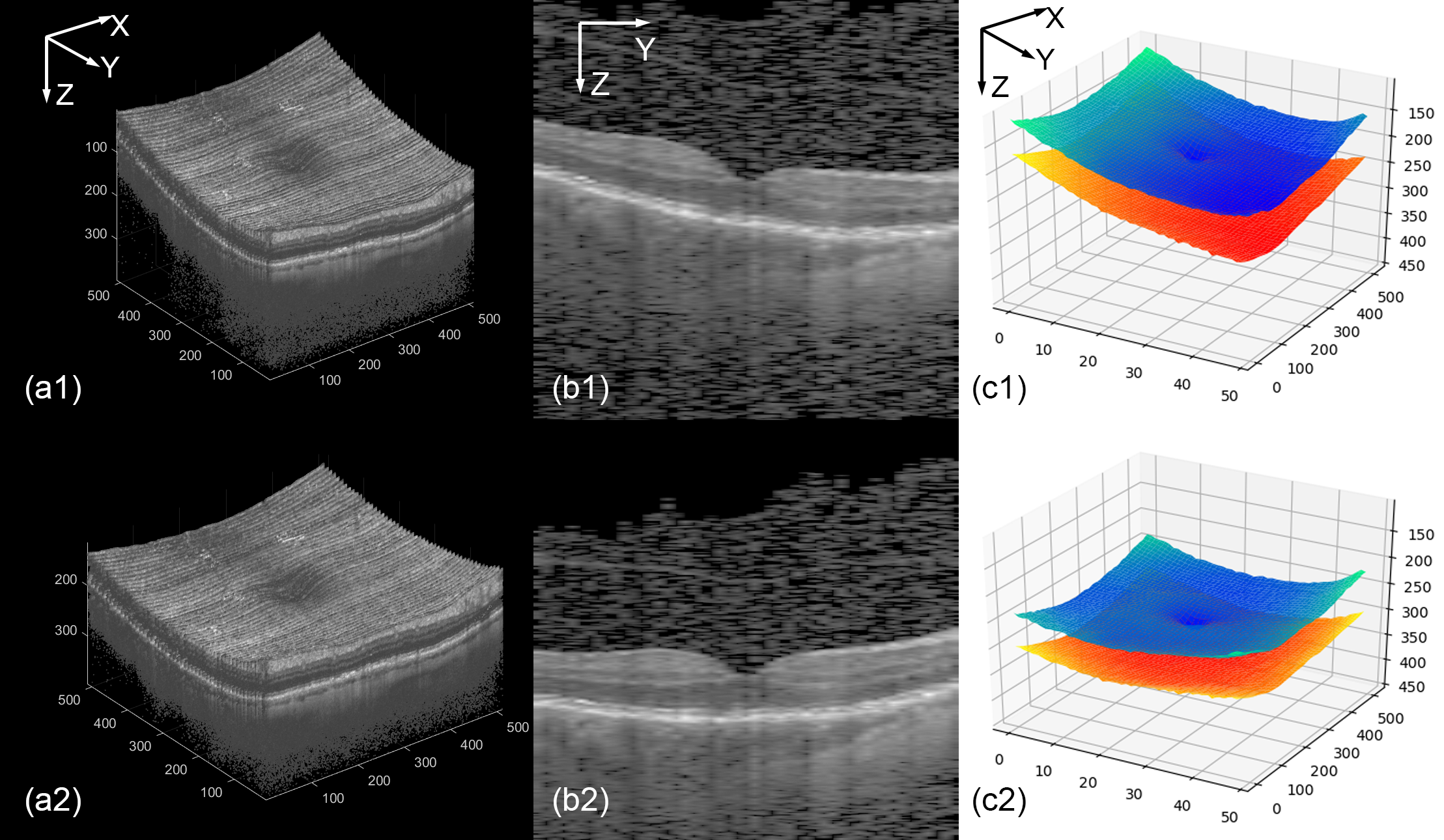}
    \caption{Tilt correction on motion corrected OCT volume. Row (1) shows the original OCT and row (2) shows the tilt corrected result. The 3D OCT volume, the cross-sectional B-scan, and ILM and RPE surfaces are presented in the columns (a), (b), and (c), respectively.}
    \label{fig:tilt_correction}
\end{figure}

We apply a tilt correction algorithm after axial motion correction for all methods at inference time, which leads to better visualization of OCT volume as illustrated in Fig.~\ref{fig:tilt_correction}. The motion corrected OCT volumes before and after tilt correction are shown in Fig.~\ref{fig:tilt_correction} row (1) and (2), respectively. 
The four corners of the RPE segmentation boundary in sub-image (c1) are mapped to the same reference plane at height $H_\mathrm{ref}=300$ by applying another displacement $\mathbf{D}_z^\mathrm{tilt}$, 
\begin{align*}
\begin{split}
    \scalemath{0.8}{\mathbf{D}_z^\mathrm{tilt}(x,y)} &= \scalemath{0.8}{H_\mathrm{ref} - \frac{1}{(W-1)(N-1)} \Big( \mathbf{B}(1,0,0)(W-1-x)(N-1-y)} \\ 
     &~+ \scalemath{0.8}{ \mathbf{B}(1,W-1,0)(N-1-y)x 
    + \mathbf{B}(1,0,N-1)(W-1-x)y } \\
    &~ \scalemath{0.8}{+ \mathbf{B}(1,W-1,N-1)xy \Big)},
\end{split}
\end{align*}
which is based on bilinear interpolation of the four corner coordinates. Finally, the tilt corrected ILM and RPE surfaces are visualized in sub-image (c2). 

\subsection{Criteria}
We evaluate the algorithms based on two criteria: the pixel-wise mean absolute error (MAE) and the smoothness of OCT scan. 
The first metric is MAE between the predicted and ground truth displacement, which measures the overall accuracy and preservation of retinal curvature, where smaller MAE indicates better performance.

\begin{equation}
    \mathrm{MAE}_z(\mathbf{D}_z;\mathbf{D}_z^\mathrm{GT}) = Z_\mathrm{norm} \underset{x,y}{\mathrm{mean}} \Big(|\mathbf{D}_z(x,y) - \mathbf{D}_z^\mathrm{GT}(x,y)| \Big).
\end{equation}
Similarly, the MAE for X motion is 
\begin{equation}
    \mathrm{MAE}_x(\mathbf{D}_x;\mathbf{D}_x^\mathrm{GT}) = X_\mathrm{norm} \underset{y}{\mathrm{mean}} \Big(|\mathbf{D}_x(y) - \mathbf{D}_x^\mathrm{GT}(y)| \Big).
\end{equation}

We also include an intensity based metric, Pearson correlation coefficient (PCC) \cite{freedman2007statistics} to measure the similarity between motion corrected OCT volume $\mathbf{V}^\mathrm{pred}$ (could be $\mathbf{V}_\mathrm{dx}$ or $\mathbf{V}_\mathrm{dz}$ depending on the experiment) and ground truth OCT volume $\mathbf{V}^\mathrm{GT}$.
The PCC, which is a commonly used metric in literature \cite{sanchez2019review}, is defined as
\begin{equation}
    \mathrm{PCC}(\mathbf{V}^\mathrm{pred}, \mathbf{V}^\mathrm{GT})  = \frac{\mathrm{cov}(\mathbf{V}^\mathrm{pred},\mathbf{V}^\mathrm{GT})}{\sigma_{\mathbf{V}^\mathrm{pred}} \sigma_{\mathbf{V}^\mathrm{GT}}},
\end{equation}
where $\sigma_{\mathbf{V}^\mathrm{pred}}$ and $\sigma_{\mathbf{V}^\mathrm{GT}}$ respectively are the standard deviations of volume $\mathbf{V}^\mathrm{pred}$ and $\mathbf{V}^\mathrm{GT}$, and the covariance can be calculated by
\begin{equation}
    \mathrm{cov}(\mathbf{V}^\mathrm{pred},\mathbf{V}^\mathrm{GT}) = \mathrm{E}[(\mathbf{V}^\mathrm{pred} - \mathrm{E}[\mathbf{V}^\mathrm{pred}])(\mathbf{V}^\mathrm{GT} - \mathrm{E}[\mathbf{V}^\mathrm{GT}])].
\end{equation}

In order to evaluate the preservation of retinal curvature, we adopt the curvature index proposed by \cite{park2019influence}. The central B-scan (X-Z plane at $Y = \mathrm{floor}(N/2)$) and the central cross sectional B-scan (Y-Z plane at $X = \mathrm{floor}(W/2)$) are taken to evaluate the X and Y curvatures, respectively. For each direction, the tilt is first corrected, and the RPE curvature is obtained by fitting a 4-th order polynomial to the segmentation boundary of the RPE using least squares. For the X direction, the polynomial coefficients $\mathbf{p}_x$ are
\begin{equation}
\scalemath{0.9}{
    \mathbf{p}_x = \underset{\mathbf{p}\in \mathbb{R}^{5\times 1}}{\arg\min} \sum_{x=0}^{W-1} \left|\mathbf{B}\left(1,x,\mathrm{floor}\Big(\frac{N}{2}\Big)\right) - [1,~ x,~ x^2,~ x^3,~ x^4] \mathbf{p} \right|^2,    
    }
\end{equation}
and similarly for the Y direction,
\begin{equation}
\scalemath{0.9}{
    \mathbf{p}_y = \underset{\mathbf{p}\in \mathbb{R}^{5\times 1}}{\arg\min} \sum_{y=0}^{N-1} \left|\mathbf{B}\left(1,\mathrm{floor}\Big(\frac{W}{2}\Big),y\right) - [1,~ y,~ y^2,~ y^3,~ y^4] \mathbf{p} \right|^2.    
    }
\end{equation}
Then the curvature index $\mathrm{Curv}$ is defined as the ratio between length of the RPE curve and the distance of a straight line after converting pixels to millimeters \cite{park2019influence}. Denoting the OCT resolution in millimeters with $L_z, L_x, L_y$, {(in our case $1.9 \times 5.8 \times 5.8$ (mm$^3$) as mentioned in subsection A.)}
\begin{equation}
    \mathrm{Curv}_x = \frac{ L_z \ca{\cdot} \mathrm{length}([1,~ x,~ x^2,~ x^3,~ x^4]\mathbf{p}_x) }{H \cdot L_x},
\end{equation}
\begin{equation}
    \mathrm{Curv}_y = \frac{ L_z \ca{\cdot} \mathrm{length}([1,~ y,~ y^2,~ y^3,~ y^4]\mathbf{p}_y) }{H \cdot L_y}.
\end{equation}
Then the distortion of curvature $\mathrm{Dist}$ to each direction is defined as the L1 difference between the curvature index of ground truth and predicted OCT volumes,
\begin{equation}
    \mathrm{Dist}_t = | \mathrm{Curv}_t^\mathrm{pred} - \mathrm{Curv}_t^\mathrm{GT} |,~~ t\in\{x,y\}. \label{eq:dist_x}
\end{equation}

As observed in \cite{park2019influence}, the X and Y directional curvature should be relatively similar. When ground truth Y curvature is not available, we evaluate L1 error between Y curvature $\mathrm{Curv}_y$ and the ground truth (same as the input) X curvature $\mathrm{Curv}_x^\mathrm{GT}$
\begin{equation}
    \mathrm{Dist}_{xy} = | \mathrm{Curv}_y^\mathrm{pred} - \mathrm{Curv}_x^\mathrm{GT} |. \label{eq:dist_xy}
\end{equation}
Finally, we use the Dice coefficient between the corrected and ground truth vessel segmentation map in the C-scan to evaluate the performance of X motion correction. Denoting the binary segmentation maps as $\mathbf{S}_1$ and $\mathbf{S}_2$, the Dice coefficient can be obtained by
\begin{equation}
    \mathrm{Dice}(\mathbf{S}_1,\mathbf{S}_2) = \frac{2\times \sum (\mathbf{S}_1 \odot \mathbf{S}_2)} {\sum\mathbf{S}_1 + \sum\mathbf{S}_2}.
    \label{eq:dice}
\end{equation}
The Dice coefficient is a value between [0,1], and higher value shows higher overlapping ratio with ground truth.
 

\subsection{Implementation}
The proposed networks are implemented in PyTorch. 
For Z motion correction, both the baseline network and the network with segmentation input have 484K model parameters. Reflected padding is used for convolutions and dropouts with $p=0.2$ is applied on every resolution level. We set $Z_\mathrm{norm} = 10$ for the normalization factor. 
The models are trained with Adam optimizer with weight decay $10^{-3}$ and batch size 4. An initial learning rate of $10^{-3}$ and exponential decay with momentum 0.995 are set to train the network for 500 max epochs and $\lambda_\mathrm{disp}=1$, $\lambda_\mathrm{smooth}=0.5$ for loss function. 
The best model, which is selected based on the lowest validation loss, is applied for the test set. 

For the X motion correction, the vessel segmentation sub-network is first trained for 500 max epochs, using Adam optimizer with weight decay $10^{-3}$, batch size 4, initial learning rate $10^{-3}$, and exponential decay with momentum 0.99. The best model selected on the validation set achieves an average segmentation accuracy of 96.05\% and Dice coefficient of 0.4776. 
After freezing the model parameters in the vessel segmentation network, the X motion prediction sub-network is trained for 1000 max epochs, using Adam optimizer with weight decay $10^{-3}$, batch size 4, and learning rate $10^{-4}$. 
We set $X_\mathrm{norm} = W/512$ for the normalization factor depending on the B-scan resolution, and use $p=0.4$ for dropouts. 
Standard data augmentation techniques including random cropping and flipping are included during training. Finally, the best model is selected on the validation set. At inference time, we convert the predicted displacement vector to integer pixels to avoid interpolation within fast B-scans.

We compare the proposed method with several other methods, which only take a single OCT volume to correct motion. Since the method in \cite{potsaid2008ultrahigh} requires orthogonal OCT volume pairs to obtain ground truth, it is excluded in the comparison. Two comparison methods \cite{montuoro2014motion, antony2011automated} are implemented in Python. The axial correction step in \cite{fu2016eye} is implemented in MATLAB based on the original authors' implementation of saliency detection \cite{fu2013cluster}.

\begin{figure*}[htb]
    \centering
    \includegraphics[width = 0.95\linewidth]{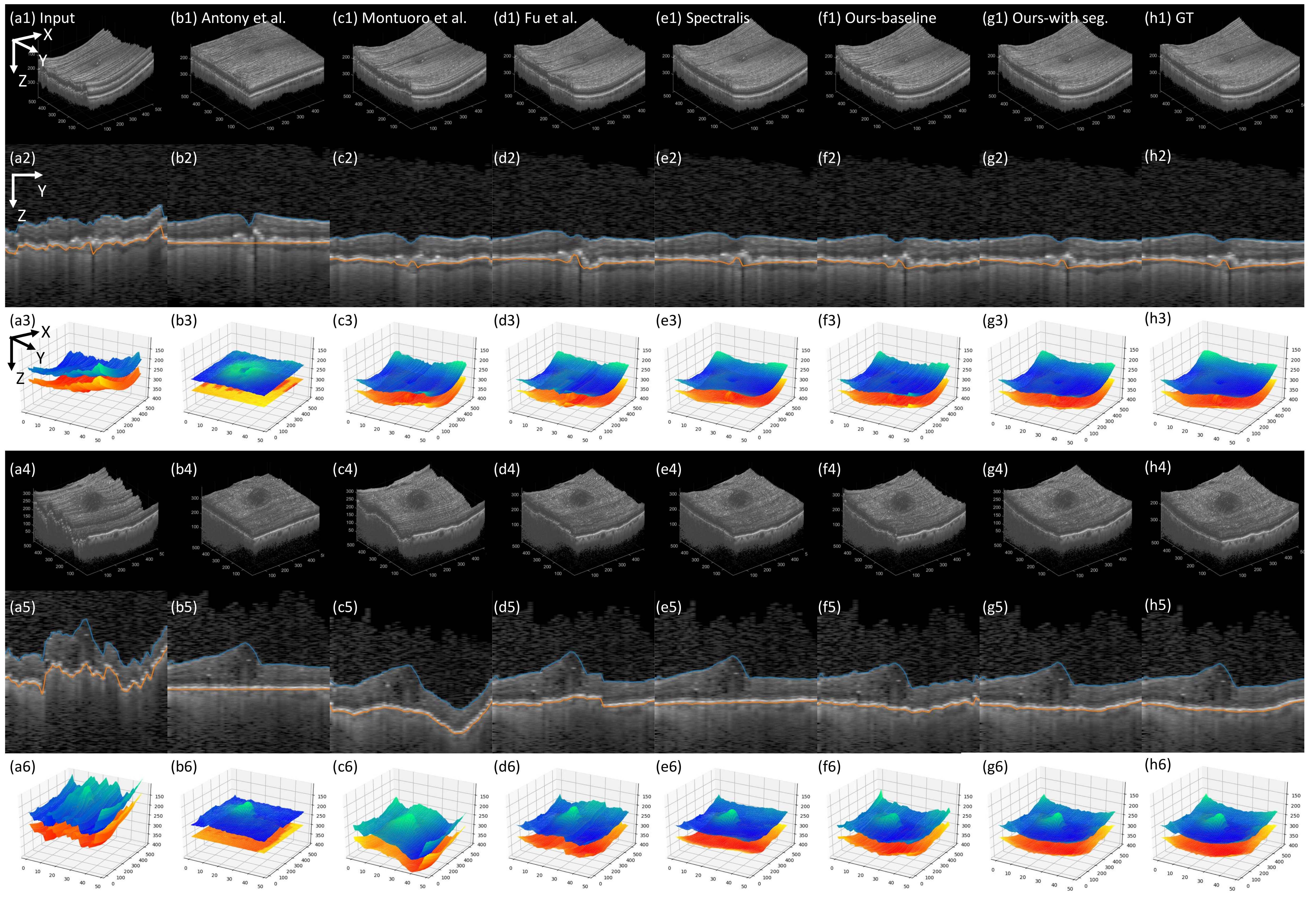}
    \caption{Qualitative result of different Z motion correction methods on the test set. Row (1-3) and (4-6) show two examples. Row (1,4) show the 3D volume, row (2,5) show the cross-sectional B-scans with segmentation boundaries of RPE and ILM, and row (3,6) show the segmentation boundaries of RPE and ILM in 3D. }
    \label{fig:moco_z}
\end{figure*}

\subsection{Evaluation on Test Dataset}
In the first experiment, we evaluate the axial and coronal motion correction methods on the testset with 37 volumes.
The qualitative results for axial motion correction using different methods are shown in Fig. \ref{fig:moco_z}. Rows (1-3) show the experimental results of one example OCT volume with moderate motion, while rows (4-6) show another example with larger motion. Rows (1,4) show the 3D volumes and rows (2,5) show the cross-sectional B-scans on the Y-Z plane with segmentation boundaries of RPE and ILM where they are overlaid onto the image in blue and orange lines, respectively, and Gamma correction at 2.2 is applied for visualization. Rows (3,6) show the segmentation boundaries of RPE and ILM after correction in 3D.

The input OCT volumes are visualized in column (a), and column (b) shows the method by Antony et al. \cite{antony2011automated}, which flattens the RPE and results in errors when the disease alters RPE in (b2). 
The method proposed by Montuoro et al. \cite{montuoro2014motion} of column (c) is able to correct axial motion without flattening the retina, but it results in unnatural curvature at large motion in (c2) and (c4) that is not similar to the ground truth in column (h).
The results of Fu et al. \cite{fu2016eye} in column (d) are smooth in most B-scans, but the errors lead to abrupt discontinuities.
The software correction of Spectralis system \cite{teussink2019spectralis} in column (e) can effectively smooth the motion, whereas the recovered curvature is different from the ground truth obtained from orthogonal OCT pairs.
Finally, our proposed baseline network in column (f) can reduce the motion artifacts in the input volume with some residual motion, while the network with segmentation input in column (g) yields smoother correction result while recovering the overall curvature.

\begin{table*}[htb]
    \caption{Quantitative result of different axial (Z) motion correction on the test set.}
    \centering
    \begin{tabular}{@{}l@{\hskip6pt}l@{\hskip6pt}c@{\hskip6pt}c@{\hskip6pt}c@{\hskip6pt}c@{\hskip6pt}c@{\hskip6pt}c@{\hskip6pt}c@{}}
    \toprule
        & & \multicolumn{7}{c}{Z correction only} \\
        \cmidrule(lr){3-9}
        \multicolumn{2}{c}{Method} & $\mathrm{MAE_z}$ & $\mathrm{PCC}(\mathbf{V}_\mathrm{dz},\mathbf{V}_\mathrm{GT})$ & $\mathrm{Curv}_x$ & $\mathrm{Dist}_x$ & $\mathrm{Curv}_y$ & $\mathrm{Dist}_y$ & $\mathrm{Dist}_{xy}$  \\
    \midrule 
        \multicolumn{2}{@{}l}{Before correction}
        & 22.742 (20.703)
        & 0.5505 (0.149)
        & 1.0020 (0.002)
        & -
        & -
        & -
        & -
        \\
        
        \multicolumn{2}{@{}l}{Ground truth}
        & -
        & 1.0000 (0.000)
        & 1.0020 (0.002)
        & -
        & 1.0013 (0.002)
        & -
        & 0.0010 (0.001)
        \\
    \midrule 
        \multicolumn{2}{@{}l}{Antony et al. \cite{antony2011automated}}
        & 40.530 (17.118)
        & 0.3517 (0.114)
        & 1.0000 (0.000)
        & 0.0020 (0.002)
        & 1.0000 (0.000)
        & 0.0012 (0.002)
        & 0.0020 (0.002)
        \\
        
        \multicolumn{2}{@{}l}{Montuoro et al. \cite{montuoro2014motion}}
        & 20.543 (19.853)
        & 0.5666 (0.152)
        & \textbf{1.0020} (0.002)
        & \textbf{0.0000} (0.000)
        & 1.0051 (0.008)
        & 0.0040 (0.007)
        & 0.0044 (0.007)
        \\
        
        \multicolumn{2}{@{}l}{Fu et al. \cite{fu2016eye}}
        & 20.963 (18.394)
        & 0.5605 (0.140)
        & \textbf{1.0020} (0.002)
        & \textbf{0.0000} (0.000)
        & 1.0048 (0.007)
        & 0.0043 (0.007)
        & 0.0042 (0.006)
        \\
        
        \multicolumn{2}{@{}l}{Spectralis \cite{teussink2019spectralis}}
        & 15.253 (9.746)
        & 0.6000 (0.140)
        & \textbf{1.0020} (0.002)
        & \textbf{0.0000} (0.000)
        & 1.0003 (0.001)
        & 0.0010 (0.001)
        & 0.0017 (0.002)
        \\
        
    \midrule
        \multirow{2}{*}{Ours (baseline)} 
        & -
        & 13.948 (10.833)
        & 0.6185 (0.112)
        & 1.0017 (0.002)
        & 0.0007 (0.001)
        & \textbf{1.0013} (0.002)
        & 0.0009 (0.001)
        & 0.0013 (0.001)
        \\
        
        & LS
        & 12.720 (9.504)
        & 0.6388 (0.114)
        & \textbf{1.0020} (0.002)
        & \textbf{0.0000} (0.000)
        & 1.0016 (0.002)
        & 0.0009 (0.001)
        & \textbf{0.0012} (0.001)
        \\
        
       \multirow{2}{*}{Ours (with seg.)}
        & -
        & 9.102 (7.972)
        & 0.7150 (0.097)
        & 1.0012 (0.001)
        & 0.0009 (0.002)
        & 1.0009 (0.001)
        & 0.0008 (0.001)
        & 0.0015 (0.002)
        \\
        
        & LS
        & \textbf{8.512} (8.211)
        & \textbf{0.7277} (0.111)
        & \textbf{1.0020} (0.002)
        & \textbf{0.0000} (0.000)
        & 1.0008 (0.001)
        & \textbf{0.0006} (0.001)
        & 0.0014 (0.002)
        \\
        
    \bottomrule
    \end{tabular}
    \label{tab:moco_z}
\end{table*}

The quantitative results of different Z motion correction are shown in Table \ref{tab:moco_z}, where each entry shows the mean and standard deviation value. We first evaluate the input OCT volumes before correction as baseline, and we also compute the curvature for ground truth. 
As the method by Antony et al. \cite{antony2011automated} flattens the retina, it achieves a curvature index of 1.0000 for both X and Y direction, and it is the only method that distorts the curvature of the X direction ($\mathrm{Dist_x}=0.0020$). It also yields larger MAE than the input (no correction) by removing the retinal curvature. The method of Montuoro et al. \cite{montuoro2014motion} and Fu et al. \cite{fu2016eye} reduce the MAE and improve PCC compared with the input by a small margin, but their Y curvature are still very different from the ground truth ($\mathrm{Dist_y}=0.0040$, $\mathrm{Dist_y}=0.0043$ respectively). The software correction result of Spectralis \cite{teussink2019spectralis} produces a MAE of 15.253 and PCC of 0.6000, and the Y curvature is flatter compared with ground truth. Overall, our method with segmentation and least squares (denoted by LS) post-processing achieves the lowest MAE at 8.512 pixels and the highest PCC at 0.7277, and the Y distortion $\mathrm{Dist_y}$  is the smallest. Our baseline network without segmentation input achieves a higher MAE and lower PCC compared with the proposed network with segmentation, ranking as the second method. The average Y curvature of our baseline network is most similar to ground truth, but the average distortion is larger compared with the network with segmentation input.

We also include an ablation study of the post-processing step, where we evaluate the raw network output displacement and the post-processed result with least squares (LS) in Table \ref{tab:moco_z}. The results show that LS post-processing lowers the MAE of both the baseline network from 13.948 to 12.720, and the network with segmentation input from 9.102 to 8.512. The PCC of the two networks increase compared to the ground truth, demonstrating that LS post-processing also improves the smoothness of corrected volume.


\begin{table}[htb]
    \caption{Evaluation of different X motion correction networks on the test set.}
    \centering
    \fontsize{7.8}{9}\selectfont
    \begin{tabular}{lc@{\hskip6pt}c@{\hskip6pt}c@{\hskip6pt}c@{\hskip6pt}c@{\hskip6pt}c@{\hskip6pt}c@{\hskip6pt}c@{}}
    \toprule
        & \multicolumn{3}{c}{GT Z correction + X correction only} \\
        \cmidrule(lr){2-4}
        Method & MAE$_x$ & PCC & Dice\\
    \midrule 
        No correction
        & 0.7484 (0.449) & 0.9714 (0.019) & 0.9205 (0.051)
        \\
        
        Ground truth
        & - & 1.0000 (0.000) & 1.0000 (0.000)
        \\
        
    \midrule 
    
        Antony et al. \cite{antony2011automated}
        & 0.7484 (0.449) & 0.9714 (0.019) & 0.9205 (0.051)
        \\
        
        Montuoro et al. \cite{montuoro2014motion}
        & 1.6152 (0.460) & 0.9506 (0.015) & 0.8191 (0.056)
        \\
        
        Fu et al. \cite{fu2016eye}
        & 0.8185 (0.438) & 0.9693 (0.019) & 0.9127 (0.049)
        \\
        
        Spectralis \cite{teussink2019spectralis}
        & - & - & -
        \\
        
    \midrule 
    
        Our X-network 
        & \textbf{0.7406} (0.437) & \textbf{0.9716} (0.019) &\textbf{0.9218} (0.048)
        \\
        
    \bottomrule
    \end{tabular}
    \label{tab:moco_x}
\end{table}

\begin{figure*}[htb]
    \centering
    \includegraphics[width = 0.9\linewidth]{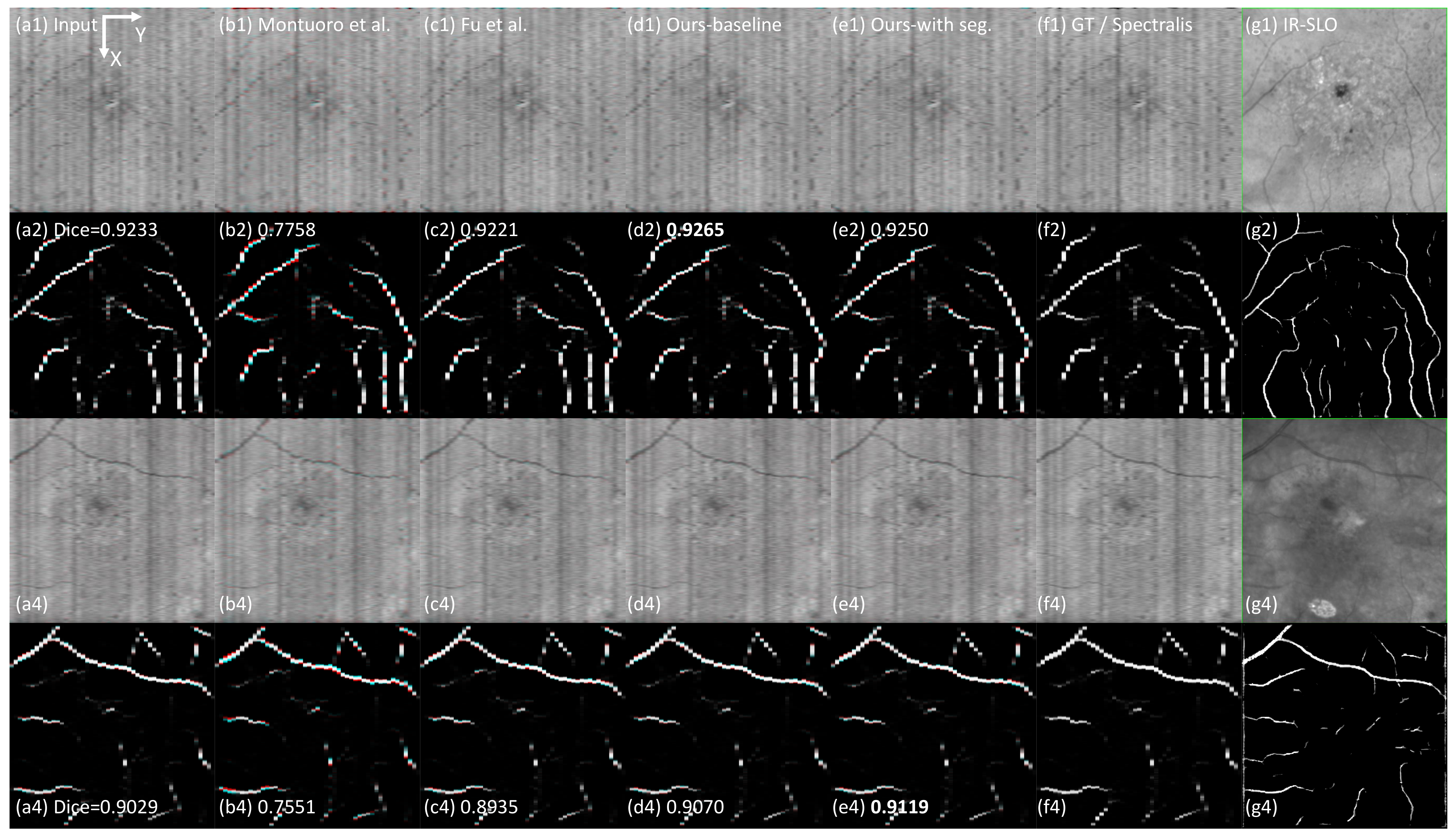}
    \caption{Qualitative result of different X motion correction methods on the test set. Row (1,3) show the overlay of C-scan of corrected OCT volumes and ground truth, and row (2,4) show the cross-sectional B-scans. The ground truth is shown in the red channel, and the corrected result is shown in cyan. Reference IR SLO images and their segmentation maps are shown in column (g). Please zoom in to see details.}
    \label{fig:moco_x}
\end{figure*}

The qualitative results of X motion correction using different methods are shown in Fig. \ref{fig:moco_x}. The en-face X-Y view of the two examples (same as Fig. \ref{fig:moco_z}) are shown in rows (1-2) and (2-4). The en-face projection C-scans shown in rows (1,3) are obtained by averaging from the ELM to RPE layer \cite{niemeijer2008vessel}, while the vessel segmentation is the intermediate output of our proposed X motion correction network. Since the corrected result is shown in cyan and the ground truth is assigned in the red channel, the overlapping region appears in gray. We also note the Dice coefficient of the overlaid segmentation maps in Fig. \ref{fig:moco_x}. The IR SLO images and their vessel segmentation maps \cite{wang2021tip} are presented in column (g) as reference for the true shape of vessels. Since the method proposed by Antony et al. \cite{antony2011automated} only includes Z motion correction, it is excluded in this comparison. The methods by Montuoro et al. \cite{montuoro2014motion} and Fu et al. \cite{fu2016eye} fail to reduce the x motion such that the Dice values of the methods are even lower than one of input (No correction). Our proposed method with both the baseline network and the network with segmentation  are able to increase the Dice coefficient and reduce the error.

The quantitative results of X motion correction is shown in Table \ref{tab:moco_x}. 
In order to evaluate X motion correction performance alone without the influence of different Z motion correction methods, we include an ablation study using ground truth Z correction input for the X motion correction step in each method. The amplitude of X motion is very small compared with Z motion in this ablation study, however, it can still be observed that only our method reduces the X motion in the input volume, achieving the lowest MAE and highest Dice coefficient. The results also demonstrate that the methods of Montuoro et al. \cite{montuoro2014motion} and Fu et al. \cite{fu2016eye} fail motion correction, reflected by higher MAE and lower Dice coefficient compared with the input (No correction).

Finally, we analyze the performance of joint Z and X motion correction algorithms based on diseases in Table \ref{tab:disease_hv}. We first test the performance with joint Z and X motion correction for each method, denoted by ``All data" in Table \ref{tab:disease_hv}. The test set is classified into the following categories: wet and dry Age-related Macular Degeneration (AMD), Diabetes, other diseases (foveal contour distortion), and normal. Overall, our proposed methods achieve the best motion correction performance for all diseases, shown by lower MAE in both Z and X direction, higher PCC, higher Dice coefficient compared with ground truth. Our proposed methods also achieve the best curvature preservation for each type of disease, demonstrated by Curv$_x$ and Curv$_y$ closer to the ground truth and smaller Dist$_x$, Dist$_y$, and Dist$_{xy}$ values.

\subsection{Evaluation on Dataset with Various Resolutions} 
In the second experiment, we test the performance of the Z and X motion correction methods on another dataset with only horizontal scan OCT volumes and different resolutions.
For X motion correction, we still utilize the Spectralis software correction result as ground truth. The MAE for X displacement and Dice coefficient between ground truth and corrected segmentation maps are evaluated. 
Since we cannot obtain axial correction ground truth via orthogonal registration from the single volume dataset, we only evaluate the distortion of curvature to the X and Y direction. 
Due to lack of ground truth, we evaluate the curvature distortion to the Y direction based on $\mathrm{Dist}_{xy}$ in eq. (\ref{eq:dist_xy}). Note that we derive the curvature distortion to the X direction $\mathrm{Dist}_{x}$ in the same way as the previous experiment using eq. (\ref{eq:dist_x}).

In Table \ref{tab:curvature_new}, we compare the evaluation results of joint Z and X motion correction on the whole dataset of 106 images, as well as on the subset of each resolution. 
For evaluation of X motion correction result on the whole dataset, our proposed methods can reduce the MAE and increase the Dice coefficient compared with the input without any correction, and the baseline network yields a slightly better result. The methods proposed by Montuoro et al. \cite{montuoro2014motion} and Fu et al. \cite{fu2016eye} on the other hand increase the MAE and decrease the Dice coefficient compared with the input, which introduce larger X motion artifacts. 
When analyzing for each resolution, our proposed method still achieves the best X motion correction performance for most resolutions, except on the resolution of $496\times 1024 \times 49$. Fu et al. \cite{fu2016eye} achieves the best performance for this resolution, while our methods remain the same as input. This may be due to the lack of data, as there are only 2 examples of this resolution. 
For Z motion correction, the numeric range of curvature index is consistent with our previous dataset. While the average Y curvature of our baseline network is more similar to the X curvature reference for all resolutions except the resolution of $496\times 512 \times 25$, our network with segmentation input achieves smaller average Y distortion for all resolutions except $496\times 1024 \times 49$. Overall, our proposed methods achieve the best preservation of retinal curvature.




\section{Conclusion}
\label{sec:conclusion}
In this paper, we proposed the deep learning neural networks for axial and coronal OCT motion correction using only a single OCT volume. The proposed baseline axial motion correction network adopts a modified U-Net architecture to predict a Z displacement map, which can be enhanced with the OCT layer segmentation. We also proposed a coronal motion network, which includes a vessel segmentation sub-network and an X motion prediction sub-network.
The experimental results show that the proposed method is able to correct axial and coronal motion while recovering the retinal curvature and achieving significant improvement compared to the conventional methods even with disease or large motion artifacts at various resolutions. 

In future work, we can generalize the proposed method to different OCT systems and investigate the possibility of training with simulation data to reduce dependency on large dataset with ground truth. The proposed method will be integrated to better display and visualization of 3D OCT scans and benefit subsequent processing including retinal layer segmentation and OCT-A imaging.

\section*{Acknowledgment}
This work is supported in part by the National Eye Institute under grant 1R01EY033847. The authors would like to thank Jacobs Retina Center for data collection.

\noindent\begin{table*}[htb]
    \caption{Quantitative result of different motion correction methods on the test set with different diseases.}
    \centering
    \fontsize{7.8}{9}\selectfont
    \begin{tabular*}{\linewidth}{@{}l@{\hskip5pt}l@{\hskip5pt}c@{\hskip5pt}c@{\hskip5pt}c@{\hskip5pt}c@{\hskip5pt}c@{\hskip5pt}c@{\hskip5pt}c@{\hskip5pt}c@{}}
    \toprule
        & & & & \multicolumn{6}{c}{Joint Z + X correction} \\
        \cmidrule(lr){5-10}
    & Metrics & Before correction & GT & Antony et al. \cite{antony2011automated} & Montuoro et al. \cite{montuoro2014motion} & Fu et al. \cite{fu2016eye} & Spectralis \cite{teussink2019spectralis} & Ours (baseline) & Ours (with seg.) \\
\midrule
    All data & MAE$_z$  & 22.742 (20.703) & - & 40.530 (17.118) & 20.543 (19.853) & 20.962 (18.394) & 15.253 (9.746) & 12.720 (9.504) & \textbf{8.5118} (8.211) \\
    & MAE$_x$ & 0.7484 (0.449) & - & - & 1.8507 (0.544) & 0.8232 (0.414) & - & \textbf{0.7428} (0.440) & 0.7401 (0.440) \\
    & PCC & 0.5505 (0.149) & - & 0.3517 (0.114) & 0.5635 (0.151) & 0.5604 (0.140) & 0.6043 (0.141) & 0.6388 (0.114) & \textbf{0.7277} (0.111) \\
    & Dice & 0.9205 (0.051) & - & - & 0.7663 (0.070) & 0.9119 (0.046) & - & \textbf{0.9214} (0.049) & 0.9219 (0.049) \\
    & Curv$_x$  & 1.0020 (0.002) & 1.0020 (0.002) & 1.0000 (0.000) & \textbf{1.0020} (0.002) & \textbf{1.0020} (0.002) & \textbf{1.0020} (0.002) & \textbf{1.0020} (0.002) & \textbf{1.0020} (0.002) \\
    & Dist$_x$  & - & - & 0.0020 (0.002) & \textbf{0.0000} (0.000) & \textbf{0.0000} (0.000) & \textbf{0.0000} (0.000) & \textbf{0.0000} (0.000) & \textbf{0.0000} (0.000) \\
    & Curv$_y$  & - & 1.0013 (0.002) & 1.0000 (0.000) & 1.0051 (0.008) & 1.0048 (0.007) & 1.0003 (0.001) & \textbf{1.0016} (0.002) & 1.0008 (0.001) \\
    & Dist$_y$  & - & - & 0.0012 (0.002) & 0.0040 (0.007) & 0.0043 (0.007) & 0.0010 (0.001) & 0.0009 (0.001) & \textbf{0.0006} (0.001) \\
    & Dist$_{xy}$  & - & - & 0.0020 (0.002) & 0.0044 (0.007) & 0.0042 (0.006) & 0.0017 (0.002) & \textbf{0.0012} (0.001) & 0.0014 (0.002) \\
\midrule
    Wet AMD & MAE$_z$  & 34.120 (22.993) & - & 48.060 (25.124) & 30.506 (21.708) & 27.466 (11.263) & 23.886 (11.221) & 20.040 (12.736) & \textbf{15.288} (12.868) \\
    & MAE$_x$ & 1.0796 (0.243) & - & - & 2.0673 (0.297) & 1.1265 (0.227) & - & 1.0735 (0.236) & \textbf{1.0673} (0.240) \\
    & PCC & 0.4866 (0.112) & - & 0.3673 (0.191) & 0.5080 (0.131) & 0.5249 (0.087) & 0.5338 (0.112) & 0.5894 (0.115) & \textbf{0.6705} (0.147) \\
    & Dice & 0.8780 (0.035) & - & - & 0.7389 (0.054) & 0.8733 (0.032) & - & 0.8792 (0.033) & \textbf{0.8803} (0.034) \\
    & Curv$_x$  & 1.0039 (0.003) & 1.0039 (0.003) & 1.0000 (0.000) & \textbf{1.0039} (0.003) & \textbf{1.0039} (0.003) & \textbf{1.0039} (0.003) & \textbf{1.0039} (0.003) & \textbf{1.0039} (0.003) \\
    & Dist$_x$  & - & - & 0.0039 (0.003) & \textbf{0.0000} (0.000) & \textbf{0.0000} (0.000) & \textbf{0.0000} (0.000) & \textbf{0.0000} (0.000) & \textbf{0.0000} (0.000) \\
    & Curv$_y$  & - & 1.0027 (0.002) & 1.0000 (0.000) & 1.0086 (0.013) & 1.0088 (0.006) & 1.0008 (0.001) & \textbf{1.0031} (0.002) & 1.0013 (0.001) \\
    & Dist$_y$  & - & - & 0.0027 (0.002) & 0.0064 (0.011) & 0.0072 (0.005) & 0.0019 (0.002) & 0.0016 (0.002) & \textbf{0.0015} (0.002) \\
    & Dist$_{xy}$  & - & - & 0.0039 (0.003) & 0.0070 (0.010) & 0.0061 (0.005) & 0.0031 (0.003) & \textbf{0.0014} (0.002) & 0.0026 (0.003) \\
\midrule
    Dry AMD & MAE$_z$  & 11.804 (6.017) & - & 45.505 (3.690) & 11.121 (6.369) & 12.051 (4.964) & 15.767 (6.265) & 10.019 (3.987) & \textbf{5.5237} (2.252) \\
    & MAE$_x$ & 0.9218 (0.146) & - & - & 2.1088 (0.340) & 0.9626 (0.116) & - & \textbf{0.9150} (0.124) & \textbf{0.9150} (0.120) \\
    & PCC & 0.6354 (0.087) & - & 0.3502 (0.010) & 0.6313 (0.101) & 0.6230 (0.071) & 0.5718 (0.106) & 0.6425 (0.101) & \textbf{0.7663} (0.069) \\
    & Dice & 0.9073 (0.027) & - & - & 0.7440 (0.070) & 0.9026 (0.026) & - & 0.9082 (0.023) & \textbf{0.9085} (0.022) \\
    & Curv$_x$  & 1.0015 (0.001) & 1.0015 (0.001) & 1.0000 (0.000) & \textbf{1.0015} (0.001) & \textbf{1.0015} (0.001) & \textbf{1.0015} (0.001) & \textbf{1.0015} (0.001) & \textbf{1.0015} (0.001) \\
    & Dist$_x$  & - & - & 0.0015 (0.001) & \textbf{0.0000} (0.000) & \textbf{0.0000} (0.000) & \textbf{0.0000} (0.000) & \textbf{0.0000} (0.000) & \textbf{0.0000} (0.000) \\
    & Curv$_y$  & - & 1.0008 (0.000) & 1.0000 (0.000) & 1.0029 (0.002) & 1.0012 (0.002) & 1.0000 (0.000) & 1.0005 (0.001) & \textbf{1.0007} (0.000) \\
    & Dist$_y$  & - & - & 0.0008 (0.000) & 0.0021 (0.002) & 0.0013 (0.002) & 0.0008 (0.000) & \textbf{0.0003} (0.000) & \textbf{0.0003} (0.000) \\
    & Dist$_{xy}$  & - & - & 0.0015 (0.001) & 0.0025 (0.002) & 0.0015 (0.001) & 0.0014 (0.001) & 0.0012 (0.001) & \textbf{0.0011} (0.001) \\
\midrule
    Diabetes & MAE$_z$  & 46.258 (29.401) & - & 36.490 (5.383) & 42.440 (31.524) & 34.825 (37.760) & 9.7601 (2.963) & 15.162 (9.132) & \textbf{7.6517} (4.250) \\
    & MAE$_x$ & 0.9694 (0.140) & - & - & 2.3214 (0.390) & 1.0867 (0.091) & - & \textbf{0.9541} (0.128) & 0.9643 (0.136) \\
    & PCC & 0.3495 (0.139) & - & 0.3025 (0.039) & 0.3687 (0.148) & 0.4743 (0.204) & 0.6248 (0.081) & 0.5715 (0.109) & \textbf{0.6948} (0.114) \\
    & Dice & 0.9028 (0.019) & - & - & 0.7489 (0.070) & 0.8909 (0.016) & - & \textbf{0.9060} (0.018) & 0.9039 (0.018) \\
    & Curv$_x$  & 1.0007 (0.001) & 1.0007 (0.001) & 1.0000 (0.000) & \textbf{1.0007} (0.001) & \textbf{1.0007} (0.001) & \textbf{1.0007} (0.001) & \textbf{1.0007} (0.001) & \textbf{1.0007} (0.001) \\
    & Dist$_x$  & - & - & 0.0007 (0.001) & \textbf{0.0000} (0.000) & \textbf{0.0000} (0.000) & \textbf{0.0000} (0.000) & \textbf{0.0000} (0.000) & \textbf{0.0000} (0.000) \\
    & Curv$_y$  & - & 1.0005 (0.000) & 1.0000 (0.000) & 1.0079 (0.007) & 1.0091 (0.013) & 1.0000 (0.000) & 1.0023 (0.002) & \textbf{1.0006} (0.000) \\
    & Dist$_y$  & - & - & 0.0005 (0.000) & 0.0075 (0.007) & 0.0090 (0.012) & 0.0004 (0.000) & 0.0018 (0.001) & \textbf{0.0003} (0.000) \\
    & Dist$_{xy}$  & - & - & 0.0007 (0.001) & 0.0073 (0.006) & 0.0086 (0.012) & 0.0006 (0.001) & 0.0019 (0.001) & \textbf{0.0004} (0.000) \\
\midrule
    Others & MAE$_z$  & 23.744 (3.940) & - & 37.313 (11.924) & 19.286 (3.597) & 19.624 (4.495) & 12.659 (6.965) & 11.626 (6.179) & \textbf{5.8620} (2.543) \\
    & MAE$_x$ & 0.9898 (0.181) & - & - & 1.8316 (0.416) & 1.0306 (0.185) & - & \textbf{0.9796} (0.184) & \textbf{0.9796} (0.184) \\
    & PCC & 0.4997 (0.029) & - & 0.3674 (0.013) & 0.5304 (0.037) & 0.5481 (0.054) & 0.6416 (0.133) & 0.6499 (0.082) & \textbf{0.7660} (0.061) \\
    & Dice & 0.8985 (0.008) & - & - & 0.7788 (0.066) & 0.8939 (0.010) & - & \textbf{0.8997} (0.007) & \textbf{0.8997} (0.007) \\
    & Curv$_x$  & 1.0010 (0.001) & 1.0010 (0.001) & 1.0000 (0.000) & \textbf{1.0010} (0.001) & \textbf{1.0010} (0.001) & \textbf{1.0010} (0.001) & \textbf{1.0010} (0.001) & \textbf{1.0010} (0.001) \\
    & Dist$_x$  & - & - & 0.0010 (0.001) & \textbf{0.0000} (0.000) & \textbf{0.0000} (0.000) & \textbf{0.0000} (0.000) & \textbf{0.0000} (0.000) & \textbf{0.0000} (0.000) \\
    & Curv$_y$  & - & 1.0008 (0.001) & 1.0000 (0.000) & 1.0082 (0.007) & 1.0043 (0.003) & 1.0000 (0.000) & 1.0014 (0.000) & \textbf{1.0008} (0.000) \\
    & Dist$_y$  & - & - & 0.0008 (0.001) & 0.0073 (0.007) & 0.0035 (0.003) & 0.0008 (0.001) & 0.0007 (0.000) & \textbf{0.0005} (0.000) \\
    & Dist$_{xy}$  & - & - & 0.0010 (0.001) & 0.0071 (0.006) & 0.0034 (0.003) & 0.0010 (0.001) & \textbf{0.0009} (0.000) & \textbf{0.0009} (0.000) \\
\midrule
    Normal & MAE$_z$  & 11.494 (8.603) & - & 34.674 (13.856) & 10.878 (8.487) & 16.219 (16.053) & 10.864 (7.048) & 7.9207 (4.214) & \textbf{5.7582} (2.287) \\
    & MAE$_x$ & 0.2712 (0.378) & - & - & 1.4260 (0.554) & 0.3807 (0.344) & - & 0.2711 (0.364) & \textbf{0.2649} (0.360) \\
    & PCC & 0.6379 (0.129) & - & 0.3507 (0.087) & 0.6451 (0.129) & 0.5891 (0.168) & 0.6556 (0.164) & 0.6925 (0.100) & \textbf{0.7524} (0.080) \\
    & Dice & 0.9715 (0.039) & - & - & 0.7990 (0.066) & 0.9580 (0.037) & - & 0.9715 (0.038) & \textbf{0.9725} (0.037) \\
    & Curv$_x$  & 1.0014 (0.001) & 1.0014 (0.001) & 1.0000 (0.000) & \textbf{1.0014} (0.001) & \textbf{1.0014} (0.001) & \textbf{1.0014} (0.001) & \textbf{1.0014} (0.001) & \textbf{1.0014} (0.001) \\
    & Dist$_x$  & - & - & 0.0014 (0.001) & \textbf{0.0000} (0.000) & \textbf{0.0000} (0.000) & \textbf{0.0000} (0.000) & \textbf{0.0000} (0.000) & \textbf{0.0000} (0.000) \\
    & Curv$_y$  & - & 1.0007 (0.001) & 1.0000 (0.000) & 1.0015 (0.002) & 1.0021 (0.006) & 1.0001 (0.000) & \textbf{1.0006} (0.001) & \textbf{1.0006} (0.000) \\
    & Dist$_y$  & - & - & 0.0007 (0.001) & 0.0009 (0.002) & 0.0023 (0.006) & 0.0007 (0.001) & 0.0005 (0.001) & \textbf{0.0003} (0.000) \\
    & Dist$_{xy}$  & - & - & 0.0014 (0.001) & 0.0015 (0.002) & 0.0029 (0.006) & 0.0013 (0.001) & \textbf{0.0009} (0.001) & 0.0010 (0.001) \\
    
    \bottomrule
    \end{tabular*}
    \label{tab:disease_hv}
\end{table*}
\FloatBarrier
\noindent\begin{table*}[!htb]
    \caption{Quantitative result of different motion correction methods on dataset with different resolutions.}
    \centering
    \fontsize{7.8}{9}\selectfont
    \begin{tabular}{@{}l@{\hskip4pt}l@{\hskip6pt}c@{\hskip6pt}c@{\hskip6pt}c@{\hskip6pt}c@{\hskip6pt}c@{\hskip6pt}c@{\hskip6pt}c@{\hskip6pt}c@{}}
    \toprule
        & & \multicolumn{6}{c}{Joint Z + X correction} \\
        \cmidrule(lr){3-9}
        Category & Metrics & Before correction & Antony et al. \cite{antony2011automated} & Montuoro et al. \cite{montuoro2014motion} & Fu et al. \cite{fu2016eye} & Spectralis \cite{teussink2019spectralis} & Ours (baseline) & Ours (with seg.) \\
    \midrule 
        All data
        & MAE$_x$
        & 0.8630 (0.227)        & -        & 8.1138 (6.219)
        & 0.9156 (0.226)        & - 
        & \textbf{0.8607} (0.226)        & 0.8619 (0.225)
        \\
        
        & Dice
        & 0.9200 (0.028)        & -        & 0.6125 (0.132)
        & 0.9158 (0.028)        & - 
        & \textbf{0.9204} (0.028)        & 0.9202 (0.028)
        \\
        
        & $\mathrm{Curv}_x$
        & 1.0012 (0.001)        & 1.0000 (0.000)        & \textbf{1.0012} (0.001)
        & \textbf{1.0012} (0.001)        & \textbf{1.0012} (0.001)  
        & \textbf{1.0012} (0.001)        & \textbf{1.0012} (0.001)
        \\
        
        & $\mathrm{Dist}_x$
        & -       & 0.0012 (0.001)        & \textbf{0.0000} (0.000)
        & \textbf{0.0000} (0.000)        & \textbf{0.0000} (0.000) 
        & \textbf{0.0000} (0.000)        & \textbf{0.0000} (0.000)
        \\
        
        & $\mathrm{Curv}_y$
        & -        & 1.0000 (0.000)        & 1.0027 (0.004)
        & 1.0027 (0.005)        & 1.0003 (0.001)
        & \textbf{1.0013} (0.001)        & 1.0007 (0.001)
        \\
        
        & $\mathrm{Dist}_{xy}$
        & -        & 0.0012 (0.001)        & 0.0025 (0.004)
        & 0.0028 (0.004)        & 0.0012 (0.002)
        & 0.0012 (0.001)        & \textbf{0.0009} (0.001)
        \\
        
    \midrule
    
        $496\times 512\times 49$
        & MAE$_x$
        & 0.8997 (0.175)        & -        & 7.6411 (5.243)
        & 0.9527 (0.180)        & -
        & \textbf{0.8988} (0.172)        & 0.8990 (0.172)
        \\
        
        & Dice
        & 0.9155 (0.024)        & -        & 0.6169 (0.132)
        & 0.9111 (0.024)        & -
        & \textbf{0.9157} (0.024)        & 0.9156 (0.024)
        \\
        
        & $\mathrm{Curv}_x$
        & 1.0012 (0.001)        & 1.0000 (0.000)        & \textbf{1.0012} (0.001)
        & \textbf{1.0012} (0.001)        & \textbf{1.0012} (0.001)
        & \textbf{1.0012} (0.001)        & \textbf{1.0012} (0.001)
        \\
        
        & $\mathrm{Dist}_x$
        & -        & 0.0012 (0.001)        & \textbf{0.0000} (0.000)
        & \textbf{0.0000} (0.000)        & \textbf{0.0000} (0.000)
        & \textbf{0.0000} (0.000)        & \textbf{0.0000} (0.000)
        \\
        
        & $\mathrm{Curv}_y$
        & -        & 1.0000 (0.000)        & 1.0024 (0.003)
        & 1.0025 (0.004)        & 1.0003 (0.001)
        & \textbf{1.0012} (0.001)        & 1.0007 (0.001)
        \\
        
        & $\mathrm{Dist}_{xy}$
        & -        & 0.0012 (0.001)        & 0.0022 (0.003)
        & 0.0025 (0.004)        & 0.0012 (0.002)
        & 0.0012 (0.002)        & \textbf{0.0009} (0.001)
        \\
        
    \midrule
    
        $496\times 512\times 25$
        & MAE$_x$
        & 0.0000 (0.000)        & -        & 10.350 (4.089)
        & 0.1067 (0.038)        & -
        & \textbf{0.0000} (0.000)        & \textbf{0.0000} (0.000)
        \\
        
        & Dice
        & 1.0000 (0.000)        & -        & 0.5449 (0.078)
        & 0.9938 (0.004)        & -
        & 1.0000 (0.000)        & 1.0000 (0.000)
        \\
        
        & $\mathrm{Curv}_x$
        & 1.0013 (0.000)        & 1.0000 (0.000)        & \textbf{1.0013} (0.000)
        & \textbf{1.0013} (0.000)        & \textbf{1.0013} (0.000)
        & \textbf{1.0013} (0.000)        & \textbf{1.0013} (0.000)
        \\
        
        & $\mathrm{Dist}_x$
        & -        & 0.0013 (0.000)        & \textbf{0.0000} (0.000)
        & \textbf{0.0000} (0.000)        & \textbf{0.0000} (0.000)
        & \textbf{0.0000} (0.000)        & \textbf{0.0000} (0.000)
        \\
        
        & $\mathrm{Curv}_y$
        & -        & 1.0000 (0.000)        & 1.0158 (0.011)
        & 1.0039 (0.004)        & 1.0016 (0.002)
        & 1.0027 (0.001)        & \textbf{1.0015} (0.001)
        \\
        
        & $\mathrm{Dist}_{xy}$
        & -        & 0.0013 (0.000)        & 0.0145 (0.011)
        & 0.0031 (0.004)        & 0.0023 (0.001)
        & 0.0014 (0.001)        & \textbf{0.0007} (0.000)
        \\
        
    \midrule
    
        $496\times 1024\times 97$
        & MAE$_x$
        & 0.7663 (0.097)        & -        & 15.238 (13.442)
        & 0.8058 (0.092)        & -
        & \textbf{0.7388} (0.110)        & 0.7577 (0.106)
        \\
        
        & Dice
        & 0.9378 (0.008)        & -       & 0.5308 (0.115)
        & 0.9346 (0.009)        & -
        & \textbf{0.9414} (0.004)        & 0.9390 (0.006)
        \\
        
        & $\mathrm{Curv}_x$
        & 1.0017 (0.002)        & 1.0000 (0.000)        & \textbf{1.0017} (0.002)
        & \textbf{1.0017} (0.002)        & \textbf{1.0017} (0.002)
        & \textbf{1.0017} (0.002)        & \textbf{1.0017} (0.002)
        \\
        
        & $\mathrm{Dist}_x$
        & -        & 0.0017 (0.002)        & \textbf{0.0000} (0.000)
        & \textbf{0.0000} (0.000)        & \textbf{0.0000} (0.000)
        & \textbf{0.0000} (0.000)        & \textbf{0.0000} (0.000)
        \\
        
        & $\mathrm{Curv}_y$
        & -        & 1.0000 (0.000)        & 1.0014 (0.002)
        & 1.0073 (0.010)        & 1.0001 (0.000)
        & \textbf{1.0011} (0.001)        & 1.0003 (0.000)
        \\
        
        & $\mathrm{Dist}_{xy}$
        & -        & 0.0017 (0.002)        & 0.0016 (0.002)
        & 0.0071 (0.009)        & 0.0016 (0.002)
        & \textbf{0.0014} (0.001)        & \textbf{0.0014} (0.002)
        \\
        
    \midrule
    
        $496\times 1024\times 49$
        & MAE$_x$
        & 0.7041 (0.092)        & -        & 5.8414 (1.168)
        & \textbf{0.6939} (0.102)        & -
        & 0.7041 (0.092)        & 0.7041 (0.092)
        \\
        
        & Dice
        & 0.9626 (0.002)        & -        & 0.7477 (0.056)
        & \textbf{0.9629} (0.002)        & -
        & 0.9626 (0.002)        & 0.9626 (0.002)
        \\
        
        & $\mathrm{Curv}_x$
        & 1.0020 (0.002)        & 1.0000 (0.000)        & \textbf{1.0020} (0.002)
        & \textbf{1.0020} (0.002)        & \textbf{1.0020} (0.002)
        & \textbf{1.0020} (0.002)        & \textbf{1.0020} (0.002)
        \\
        
        & $\mathrm{Dist}_x$
        & -        & 0.0020 (0.002)        & \textbf{0.0000} (0.000)
        & \textbf{0.0000} (0.000)        & \textbf{0.0000} (0.000)
        & \textbf{0.0000} (0.000)        & \textbf{0.0000} (0.000)
        \\
        
        & $\mathrm{Curv}_y$
        & -        & 1.0000 (0.000)        & 1.0002 (0.000)
        & 1.0007 (0.000)        & 1.0003 (0.000)
        & \textbf{1.0014} (0.001)        & 1.0008 (0.001)
        \\
        
        & $\mathrm{Dist}_{xy}$
        & -        & 0.0020 (0.002)        & 0.0018 (0.002)
        & 0.0018 (0.001)        & 0.0017 (0.002)
        & \textbf{0.0009} (0.001)        & 0.0012 (0.001)
        \\

    \bottomrule
    
    \end{tabular}
    \label{tab:curvature_new}
\end{table*}

\FloatBarrier

\bibliographystyle{ieeetr} 
\bibliography{refs}{}


\vskip -3\baselineskip plus -1fil

\begin{IEEEbiography}[{\includegraphics[width=2.5cm]{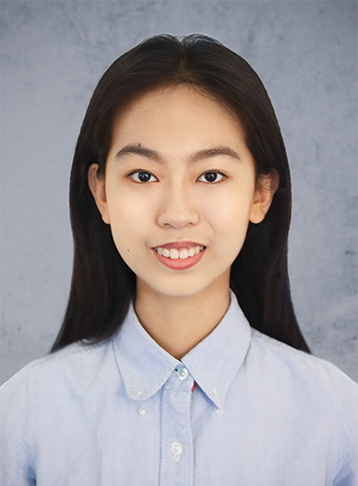}}]{Yiqian Wang}
is currently a Ph.D. candidate in the Electrical and Computer Engineering Department, University of California, San Diego. She received her B.S. degree in Electrical Engineering from Beijing Institute of Technology, Beijing, China, in 2018. Her research interests include medical image processing, signal processing, and deep learning.
\end{IEEEbiography}
\vskip -2\baselineskip plus -1fil

\begin{IEEEbiography}[{\includegraphics[width=2.5cm]{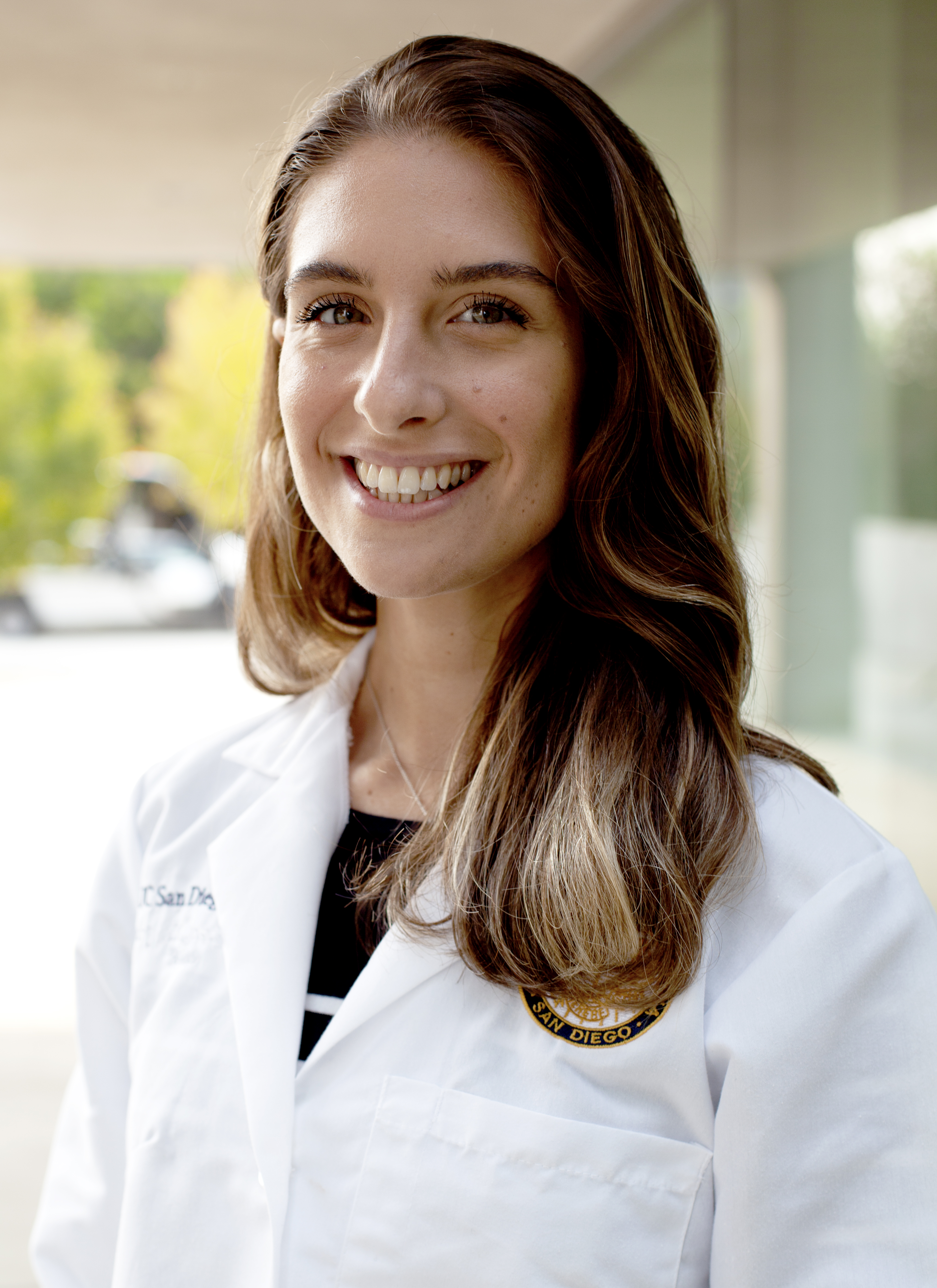}}]{Alexandra Warter}
received her M.D. degree from the University Institute of Medicine of the Hospital Italiano, Buenos Aires, Argentina, in 2015 and completed her Ophthalmology Residency at the Centro de Ojos Quilmes, Buenos Aires, Argentina, in 2020. She is currently a Clinical and Research Retina Fellow at the Jacobs Retina Center, Shiley Eye Institute, University of California San Diego. Her research and clinical interests include Macular Degeneration, Diabetic Retinopathy and imaging of these and related diseases as well as all areas of retina disorders.  She is also focused on clinical trial methodology in retina.
\end{IEEEbiography}
\vskip -2\baselineskip plus -1fil

\begin{IEEEbiography}[{\includegraphics[width=2.5cm]{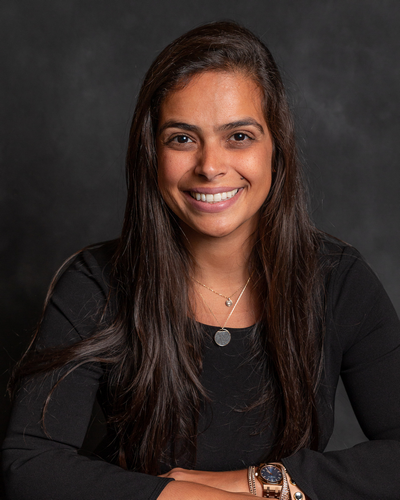}}]{Melina Cavichini}
is currently a Retina Fellow Research at Jacobs Retina Center in the Shiley Eye Institute, University of California, San Diego. She did her Vitreo Retinal Surgery Fellowship at Suel Abujamra Institute, São Paulo, Brazil in 2014, her Ophthalmology Residency at School of Medicine from ABC region, São Paulo, Brazil in 2011, and her MD degree is from University Gama Filho, Rio de Janeiro, Brazil in 2008. Her research interest include retinal diseases, the use of artificial intelligence in retina and new technologies for retina.
\end{IEEEbiography}
\vskip -2\baselineskip plus -1fil

\begin{IEEEbiography}[{\includegraphics[width=2.5cm]{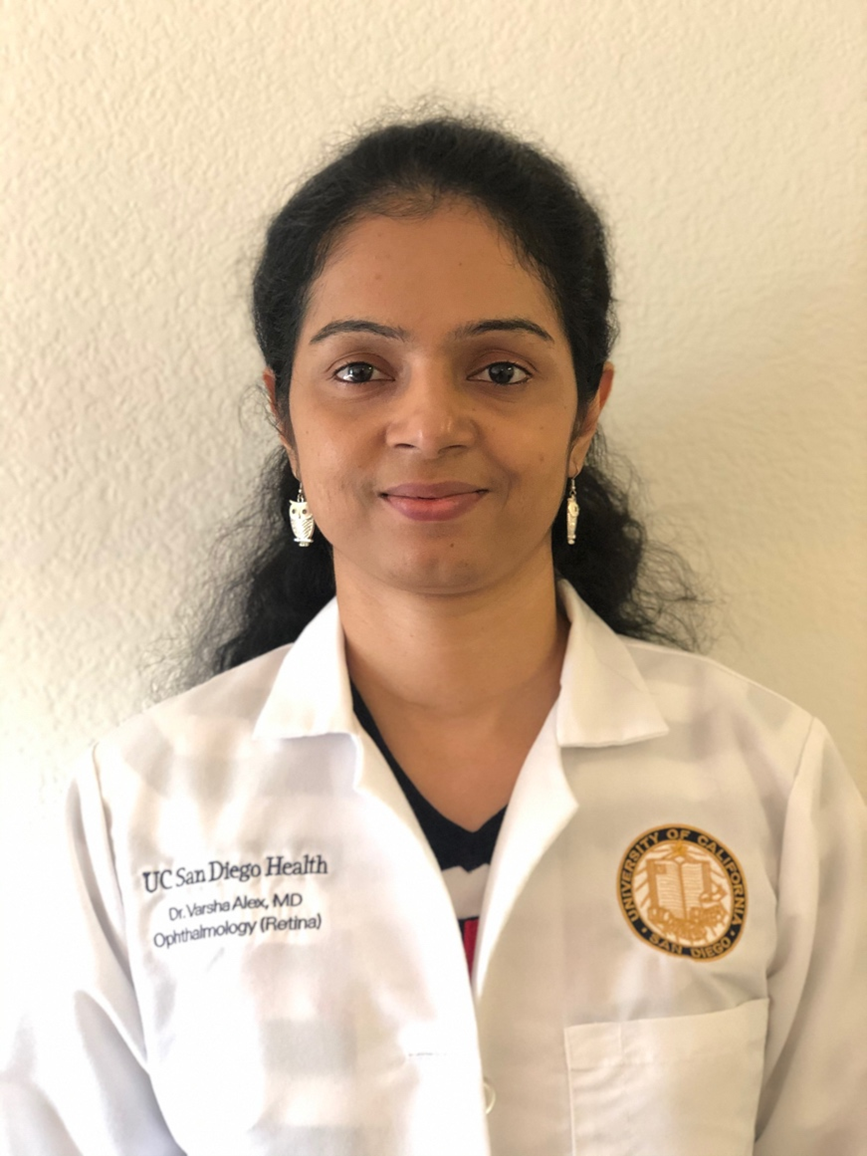}}]{Varsha Alex}
finished her undergraduate studies in Medicine and Masters in Ophthalmology from Christian Medical College, (CMC) Vellore, India, one of top medical institutes in India. After postgraduation, she joined as a retina research fellow at Shiley Eye Institute, UC San Diego. Her interests are in medical retina imaging and diagnostics and is particularly focused on translational research. She has served as consultant physician at multiple hospitals and published papers in peer reviewed journals. 
\end{IEEEbiography}
\vskip -2\baselineskip plus -1fil

\begin{IEEEbiography}[{\includegraphics[width=2.5cm]{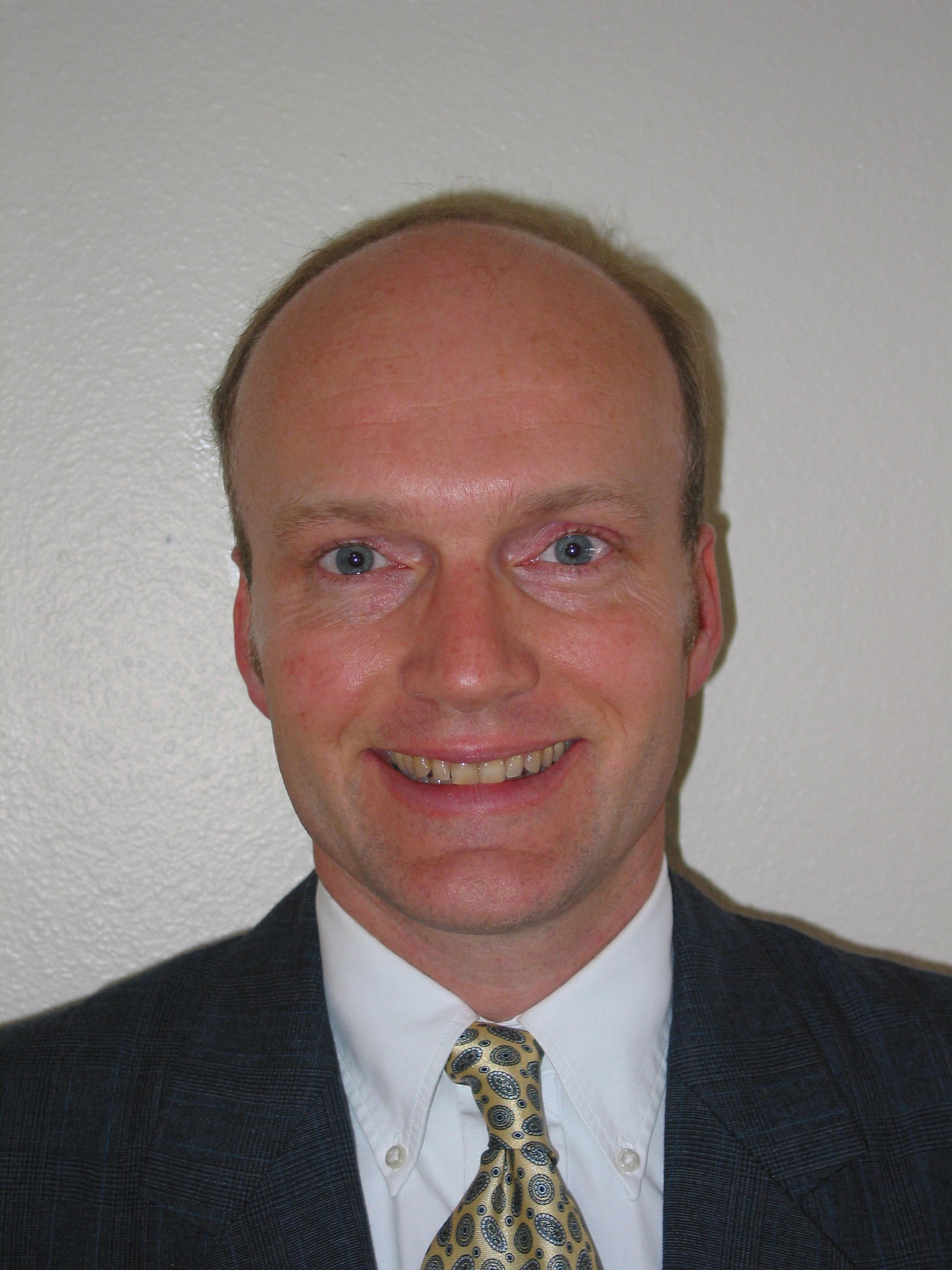}}]{Dirk-Uwe G. Bartsch}
is Associate Adjunct Professor and Co-director of the Jacobs Retina Center. Dr. Bartsch attended Technische Universitaet Darmstadt for his undergraduate degree and went on to complete his Ph.D. in bioengineering and post-doctoral fellowship at University of California, San Diego.
Dr. Bartsch’s research is focused in retinal imaging, scanning laser imaging -confocal/non-confocal, optical coherence tomography (OCT), indocyanine green and fluorescein angiography, and tomographic reconstruction of the posterior pole in patients with various retina diseases such as age-related macular degeneration, diabetes and HIV-related complications.
\end{IEEEbiography}
\vskip -2\baselineskip plus -1fil

\begin{IEEEbiography}[{\includegraphics[width=2.5cm]{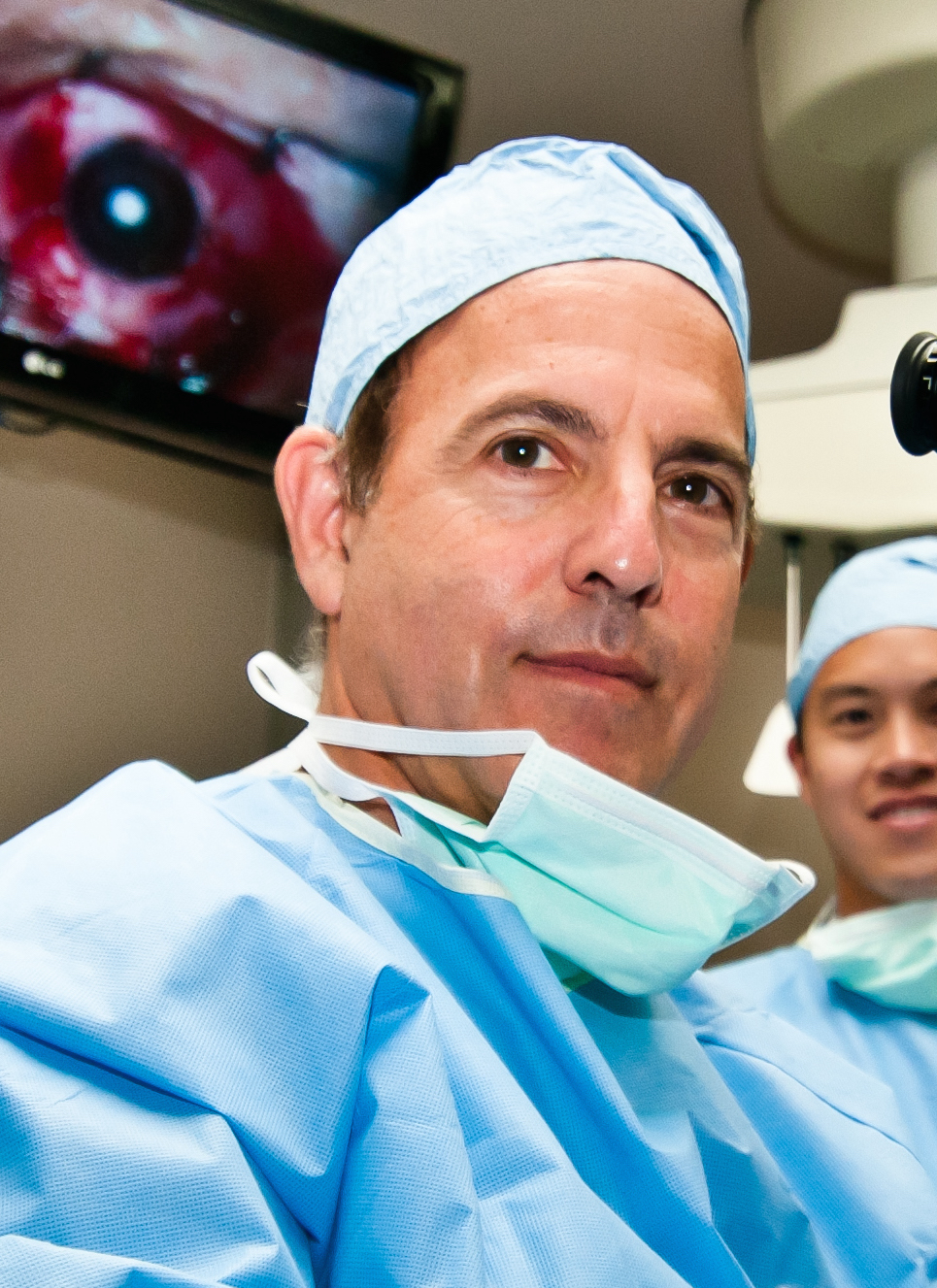}}]{William R. Freeman}
is Distinguished Professor of Ophthalmology, Director of the UCSD Jacobs Retina Center and Vice Chair of the UCSD Department of Ophthalmology.  He is a full time Retina Surgeon and also a researcher who has held NIH grants for nearly 30 years.  He works closely with imaging groups in the department of Ophthalmology as well as in the UCSD School of Engineering.  He has over 600 peer reviewed publications.
\end{IEEEbiography}
\vskip -2\baselineskip plus -1fil

\begin{IEEEbiography}[{\includegraphics[width=2.5cm]{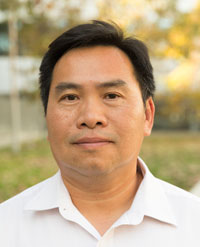}}]{Truong Q. Nguyen}
[F'05] is currently a Professor at the ECE Dept., UC San Diego. His current research interests are 3D video processing and communications and their efficient implementation.  He is the coauthor (with Prof. Gilbert Strang) of a popular textbook, Wavelets \& Filter Banks, Wellesley-Cambridge Press, 1997, and the author of several matlab-based toolboxes on image compression, electrocardiogram compression and filter bank design.  He has over 400 publications.

Prof. Nguyen  received the IEEE Transaction in Signal Processing Paper Award (Image and Multidimensional Processing area) for the paper he co-wrote with Prof. P. P. Vaidyanathan on linear-phase perfect-reconstruction filter banks (1992).  He received the NSF Career Award in 1995 and is currently the Series Editor (Digital Signal Processing) for Academic Press.  He served as  Associate Editor for the IEEE Transaction on Signal Processing 1994-96, for the Signal Processing Letters 2001-2003, for the IEEE Transaction on Circuits \& Systems from 1996-97, 2001-2004, and for the IEEE Transaction on Image Processing from 2004-2005. 
\end{IEEEbiography}
\vskip -2\baselineskip plus -1fil

\begin{IEEEbiography}[{\includegraphics[width=2.5cm]{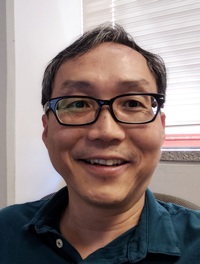}}]{Cheolhong An}
is an assistant adjunct professor at the Electrical and Computer Engineering, University of California, San Diego. Earlier, he worked at Samsung Electronics, Korea and Qualcomm, USA. He received the B.S. and M.S. degrees in electrical engineering from Pusan National University, Busan, Korea, in 1996 and 1998, respectively, and Ph.D. in Electrical and Computer Engineering in 2008. His current research is focused on the medical image processing and the real-time bio image processing. His research interests are in 2D and 3D image processing with machine learning and sensor technology.
\end{IEEEbiography}


\end{document}


\maketitle


\noindent\begin{table*}[!htb]
    \caption{Quantitative result of different motion correction methods on dataset with different resolutions.}
    \centering
    \fontsize{7.8}{9}\selectfont
    \begin{tabular}{@{}l@{\hskip4pt}l@{\hskip6pt}c@{\hskip6pt}c@{\hskip6pt}c@{\hskip6pt}c@{\hskip6pt}c@{\hskip6pt}c@{\hskip6pt}c@{\hskip6pt}c@{}}
    \toprule
        & & \multicolumn{6}{c}{Joint Z + X correction} \\
        \cmidrule(lr){3-9}
        Category & Metrics & Before correction & Antony et al. \cite{antony2011automated} & Montuoro et al. \cite{montuoro2014motion} & Fu et al. \cite{fu2016eye} & Spectralis \cite{teussink2019spectralis} & Ours (baseline) & Ours (with seg.) \\
    \midrule 
        All data
        & MAE$_x$
        & 0.8630 (0.227)        & -        & 8.1138 (6.219)
        & 0.9156 (0.226)        & - 
        & \textbf{0.8607} (0.226)        & 0.8619 (0.225)
        \\
        
        & Dice
        & 0.9200 (0.028)        & -        & 0.6125 (0.132)
        & 0.9158 (0.028)        & - 
        & \textbf{0.9204} (0.028)        & 0.9202 (0.028)
        \\
        
        & $\mathrm{Curv}_x$
        & 1.0012 (0.001)        & 1.0000 (0.000)        & \textbf{1.0012} (0.001)
        & \textbf{1.0012} (0.001)        & \textbf{1.0012} (0.001)  
        & \textbf{1.0012} (0.001)        & \textbf{1.0012} (0.001)
        \\
        
        & $\mathrm{Dist}_x$
        & -       & 0.0012 (0.001)        & \textbf{0.0000} (0.000)
        & \textbf{0.0000} (0.000)        & \textbf{0.0000} (0.000) 
        & \textbf{0.0000} (0.000)        & \textbf{0.0000} (0.000)
        \\
        
        & $\mathrm{Curv}_y$
        & -        & 1.0000 (0.000)        & 1.0027 (0.004)
        & 1.0027 (0.005)        & 1.0003 (0.001)
        & \textbf{1.0013} (0.001)        & 1.0007 (0.001)
        \\
        
        & $\mathrm{Dist}_{xy}$
        & -        & 0.0012 (0.001)        & 0.0025 (0.004)
        & 0.0028 (0.004)        & 0.0012 (0.002)
        & 0.0012 (0.001)        & \textbf{0.0009} (0.001)
        \\

\midrule
    Wet AMD& MAE$_x$ & 0.8625 (0.201) & - & 7.8896 (7.100) & 0.9074 (0.208) & - & \textbf{0.8633} (0.199) & 0.8649 (0.199) \\
    & Dice & 0.9245 (0.028) & - & 0.6461 (0.138) & 0.9201 (0.029) & - & \textbf{0.9245} (0.028) & 0.9242 (0.028) \\
    & Curv$_x$  & 1.0016 (0.002) & 1.0000 (0.000) & \textbf{1.0016} (0.002) & \textbf{1.0016} (0.002) & \textbf{1.0016} (0.002) & \textbf{1.0016} (0.002) & \textbf{1.0016} (0.002) \\
    & Dist$_x$  & - & 0.0016 (0.002) & \textbf{0.0000} (0.000) & \textbf{0.0000} (0.000) & \textbf{0.0000} (0.000) & \textbf{0.0000} (0.000) & \textbf{0.0000} (0.000) \\
    & Curv$_y$  & - & 1.0000 (0.000) & 1.0027 (0.004) & 1.0035 (0.006) & 1.0004 (0.001) & \textbf{1.0013} (0.001) & 1.0007 (0.001) \\
    & Dist$_{xy}$  & - & 0.0016 (0.002) & 0.0030 (0.004) & 0.0037 (0.005) & 0.0015 (0.002) & 0.0014 (0.002) & \textbf{0.0012} (0.002) \\
\midrule
    Dry AMD & MAE$_x$ & 0.8536 (0.246) & - & 8.0840 (5.272) & 0.9148 (0.238) & - & \textbf{0.8482} (0.245) & 0.8494 (0.244) \\
    & Dice & 0.9174 (0.028) & - & 0.5960 (0.120) & 0.9132 (0.027) & - & \textbf{0.9181} (0.028) & 0.9179 (0.027) \\
    & Curv$_x$  & 1.0011 (0.001) & 1.0000 (0.000) & \textbf{1.0011} (0.001) & \textbf{1.0011} (0.001) & \textbf{1.0011} (0.001) & \textbf{1.0011} (0.001) & \textbf{1.0011} (0.001) \\
    & Dist$_x$  & - & 0.0011 (0.001) & \textbf{0.0000} (0.000) & \textbf{0.0000} (0.000) & \textbf{0.0000} (0.000) & \textbf{0.0000} (0.000) & \textbf{0.0000} (0.000) \\
    & Curv$_y$  & - & 1.0000 (0.000) & 1.0030 (0.005) & 1.0024 (0.004) & 1.0003 (0.002) & \textbf{1.0012} (0.001) & 1.0008 (0.001) \\
    & Dist$_{xy}$  & - & 0.0011 (0.001) & 0.0025 (0.004) & 0.0024 (0.003) & 0.0012 (0.002) & 0.0011 (0.001) & \textbf{0.0007} (0.001) \\
\midrule
    Other Diseases & MAE$_x$ & 1.0490 (0.213) & - & 12.9876 (7.759) & 1.1061 (0.206) & - & \textbf{1.0408} (0.221) & \textbf{1.0408} (0.219) \\
    & Dice & 0.9046 (0.026) & - & 0.5332 (0.118) & 0.8993 (0.027) & - & \textbf{0.9062} (0.027) & 0.9058 (0.027) \\
    & Curv$_x$  & 1.0007 (0.001) & 1.0000 (0.000) & \textbf{1.0007} (0.001) & \textbf{1.0007} (0.001) & \textbf{1.0007} (0.001) & \textbf{1.0007} (0.001) & \textbf{1.0007} (0.001) \\
    & Dist$_x$  & - & 0.0007 (0.001) & \textbf{0.0000} (0.000) & \textbf{0.0000} (0.000) & \textbf{0.0000} (0.000) & \textbf{0.0000} (0.000) & \textbf{0.0000} (0.000) \\
    & Curv$_y$  & - & 1.0000 (0.000) & 1.0010 (0.001) & 1.0015 (0.002) & 1.0003 (0.000) & 1.0021 (0.003) & \textbf{1.0008} (0.001) \\
    & Dist$_{xy}$  & - & \textbf{0.0007} (0.001) & \textbf{0.0007} (0.001) & 0.0015 (0.002) & 0.0009 (0.001) & 0.0023 (0.003) & 0.0011 (0.001) \\
\midrule
    Normal & MAE$_x$ & 0.7925 (0.153) & - & 5.9179 (2.420) & 0.8231 (0.173) & - & \textbf{0.7993} (0.151) & \textbf{0.7993} (0.151) \\
    & Dice & 0.9238 (0.019) & - & 0.5801 (0.134) & 0.9207 (0.021) & - & \textbf{0.9234} (0.019) & \textbf{0.9234} (0.019) \\
    & Curv$_x$  & 1.0003 (0.000) & 1.0000 (0.000) & \textbf{1.0003} (0.000) & \textbf{1.0003} (0.000) & \textbf{1.0003} (0.000) & \textbf{1.0003} (0.000) & \textbf{1.0003} (0.000) \\
    & Dist$_x$  & - & 0.0003 (0.000) & \textbf{0.0000} (0.000) & \textbf{0.0000} (0.000) & \textbf{0.0000} (0.000) & \textbf{0.0000} (0.000) & \textbf{0.0000} (0.000) \\
    & Curv$_y$  & - & 1.0000 (0.000) & 1.0012 (0.001) & 1.0010 (0.001) & 1.0000 (0.000) & \textbf{1.0005} (0.001) & \textbf{1.0005} (0.000) \\
    & Dist$_{xy}$  & - & \textbf{0.0003} (0.000) & 0.0009 (0.001) & 0.0010 (0.001) & \textbf{0.0003} (0.000) & 0.0007 (0.000) & 0.0004 (0.000) \\ 
    \bottomrule
    
    \end{tabular}
    \label{tab:curvature_new}
\end{table*}

\begin{figure*}[htb]
    \centering
    \includegraphics[width = 1\linewidth]{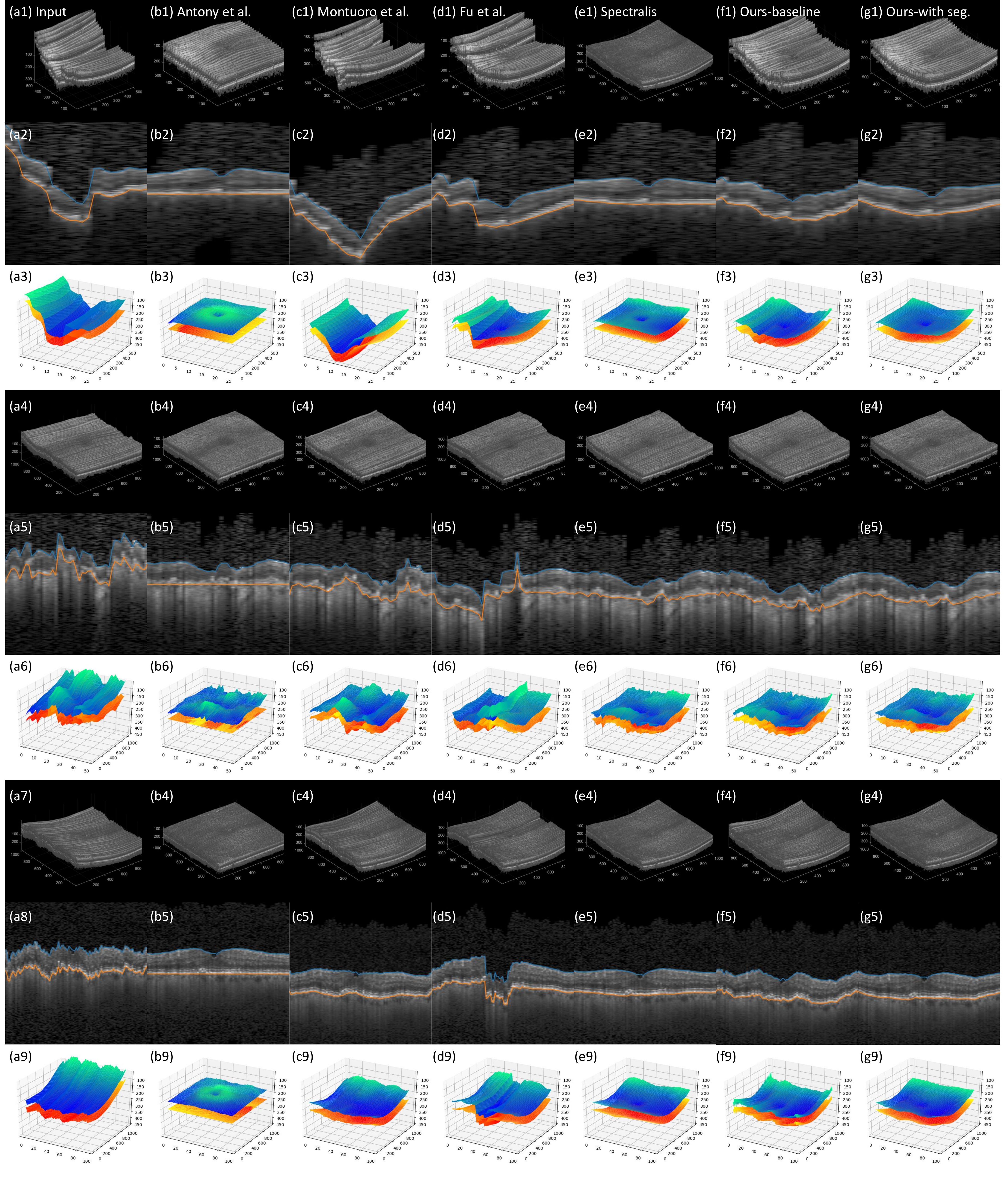}
    \caption{Qualitative result of different Z motion correction methods on different resolutions. Row (1-3), (4-6), (7-9) show three examples. Row (1,4,7) show the 3D volume, row (2,5,8) show the cross-sectional B-scans with segmentation boundaries of RPE and ILM, and row (3,6,9) show the segmentation boundaries of RPE and ILM in 3D. }
    \label{fig:moco_new}
\end{figure*}

We show the qualitative result of axial motion correction for three example OCT volumes in Fig. \ref{fig:moco_new} with resolutions $496\times 512\times 25$ (rows 1-3), $496\times 1024\times 49$ (rows 4-6), and $496\times 1024\times 97$ (rows 7-9) in a similar way as Fig. 9 in the original manuscript. 
Overall, the method proposed by Antony et al. \cite{antony2011automated} flattens the retina and is inaccurate in the second example with disease. The methods by Montuoro et al. \cite{montuoro2014motion} and Fu et al. \cite{fu2016eye} cannot effectively reduce axial motion for large motion in the first example and disease in the second example, and the method by Fu et al. also suffers from discontinuities in the third example. The Spectralis \cite{teussink2019spectralis} software correction result is visually smooth, but the retinal curvature looks flattened along the slow scanning axis. Even though our method with baseline axial motion correction network has visible residual artifacts especially when the OCT resolution is increased, our method with segmentation can still achieve desirable correction while recovering a motion-free retina with natural curvature. 
It is worth noting that even when these different resolutions are not included during training, the result demonstrates that our proposed network can generalize well to various resolutions.

\begin{figure*}[htb]
    \centering
    \includegraphics[width = 1\linewidth]{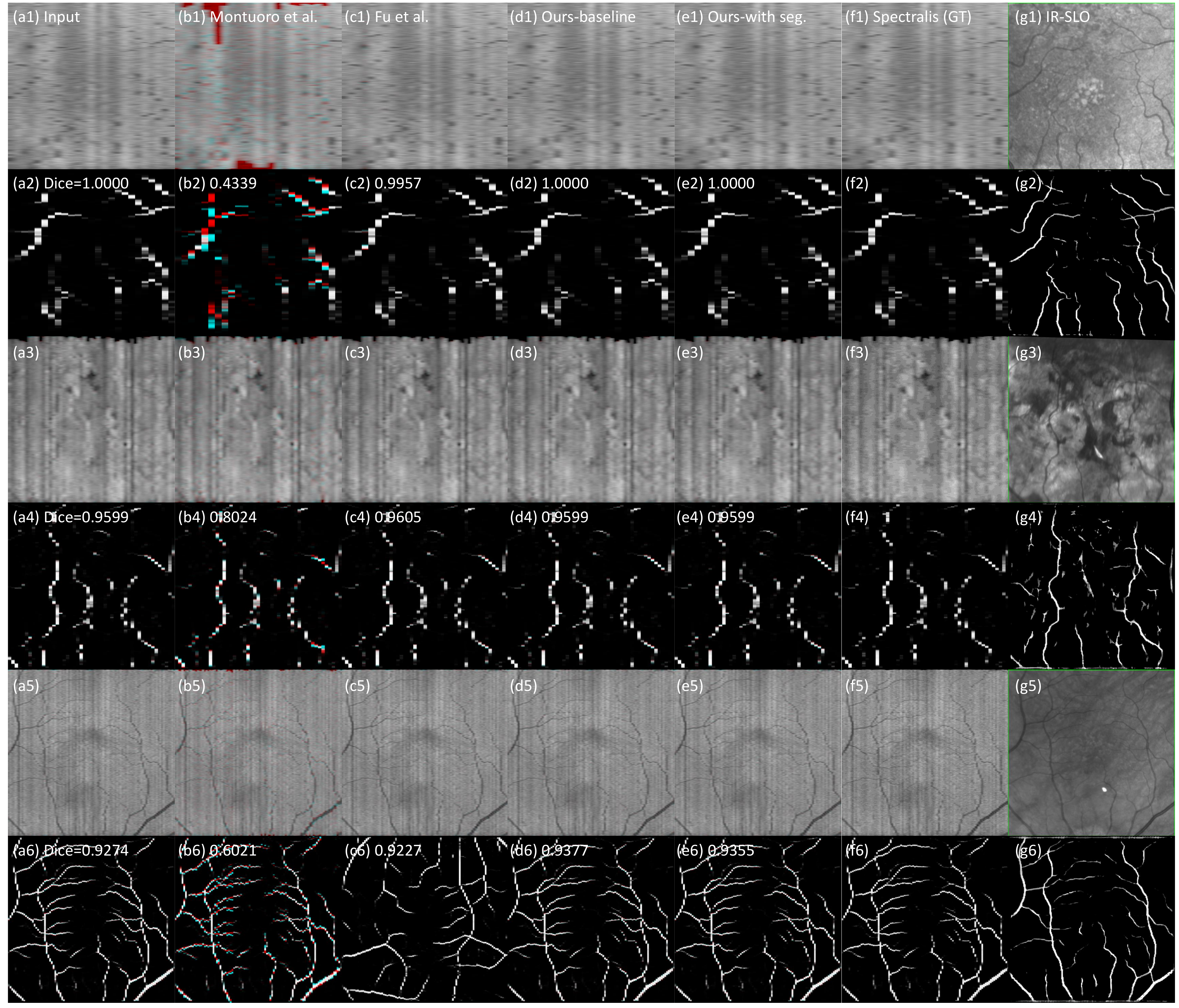}
    \caption{Qualitative result of different X motion correction methods on different resolutions. Row (1,3,5) show the overlay of C-scan of corrected OCT volumes and ground truth, and row (2,4,6) show the cross-sectional B-scans. The ground truth is shown in the red channel, and the corrected result is shown in cyan. Reference IR SLO images and their segmentation maps are shown in column (g). Please zoom in to see details.}
    \label{fig:moco_x_new}
\end{figure*}

The X motion correction performance using different methods is also compared qualitatively as shown in Fig. \ref{fig:moco_x_new}. Note that there is no X motion in the input OCT volume for the first example with resolution $496\times 512\times 25$, because the scanning speed is low enough for the hardware eye-tracker to compensate for X motion without residual error. Our proposed networks are robust to this example and predicts zero X motion, which does not causes performance regression. The other two methods however introduce undesirable artifacts. For the second example with severe disease, our method could not reduce the motion, whereas the other two methods lead to regression. For the third example of the highest resolution, our proposed networks can reduce the X motion and our method with baseline axial motion correction network yields a slightly higher Dice coefficient.

\bibliographystyle{ieeetr} 
\bibliography{refs}{}